\documentclass[useAMS,usenatbib,usegraphicx]{mn2e}
\usepackage{url}
\usepackage{hyperref}

\usepackage{threeparttable}
\usepackage{stmaryrd}

\def\aj{AJ}%
\def\apj{ApJ}%
\def\apjs{ApJS}%
\def\aap{A\&A}%
\def\mnras{MNRAS}%
\def\hi{H\,{\sc i}}

\begin{document}

\title[Merger Visibility Times in \hi]{Quantified \hi \ Morphology III: Merger Visibility Times from \hi \ in Galaxy Simulations}

\author[B.W. Holwerda et al.]{B. W. Holwerda,$^{1,2}$\thanks{E-mail:
benne.holwerda@esa.int}  N. Pirzkal,$^{3}$ T. J. Cox,$^{4}$ W. J. G. de Blok,$^{2}$ J. Weniger,$^{5}$ 
\newauthor
A. Bouchard,$^{6}$ S.-L. Blyth,$^{2}$ and K. S. van der Heyden$^{2}$, \\
$^{1}$ European Space Agency, ESTEC, Keplerlaan 1, 2200 AG, Noordwijk, the Netherlands\\
$^{2}$ Astrophysics, Cosmology and Gravity Centre ($AC \lightning GC$), \\
Astronomy Department, University of Cape Town, Private Bag X3, 7700 Rondebosch, Republic of South Africa\\
$^{3}$ Space Telescope Science Institute, Baltimore, MD 21218, USA\\
$^{4}$ Carnegie Observatories, 813 Santa Barbara Street, Pasadena, California, 91101, USA\\
$^{5}$ Institut f\"ur Astronomie, Universit\"at Wien, T\"urkenschanzstra\ss{}e 17, 1180 Wien, Austria\\
$^{6}$ Department of Physics, Rutherford Physics Building, McGill University, 3600 University Street, Montreal, Quebec, H3A 2T8, Canada}

\date{Accepted ----. Received ---- ; in original form June 2010}

\pagerange{\pageref{firstpage}--\pageref{lastpage}} \pubyear{2010}

\maketitle

\label{firstpage}

\begin{abstract}
Major mergers of disk galaxies are thought to be a substantial driver in galaxy evolution. To trace the fraction and the rate galaxies are in mergers over cosmic times, several observational techniques, including morphological selection criteria, have been developed over the last decade. We apply this morphological selection of mergers to 21 cm radio emission line (\hi) column density images of spiral galaxies in nearby surveys. In this paper, we investigate how long a 1:1 merger is visible in \hi \ from N-body simulations. 

We evaluate the merger visibility times for selection criteria based on four parameters: Concentration, Asymmetry, $M_{20}$ and the Gini parameter of second order moment of the flux distribution ($G_M$). Of three selection criteria used in the literature, one based on Concentration and $M_{20}$ works well for the \hi \ perspective with a merger time scale of 0.4 Gyr.
Of the three selection criteria defined in our previous paper, the $G_M$ performs well and cleanly selects mergers for 0.69 Gyr. The other two criteria (A-$M_{20}$ and C-$M_{20}$), 
select isolated disks as well, but perform best for face-on, gas-rich disks ($T_{mgr}\sim1$ Gyr).
The different visibility scales can be combined with the selected fractions of galaxies in any large \hi \ survey to obtain merger rates in the nearby Universe. All-sky surveys such as WALLABY with ASKAP and the Medium Deep Survey with the APETIF instrument on Westerbork are set to revolutionize our perspective on neutral hydrogen and will provide an accurate measure of the merger fraction and rate of the present epoch.
%


\end{abstract}


\section{\label{s:intro}Introduction}

The merger rate of galaxies over cosmic times is one of the big outstanding questions in the evolution of galaxies. Two recent major observational efforts 
seek to address the rate of galaxy mergers, one using close galaxy pairs \citep[e.g.,][]{Patton97,Le-Fevre00}, and one using quantified morphology \citep[e.g.,][]{CAS,Lotz04}. 
Both give a fraction of the population at a given redshift that is merging ($f_m$) which then can be converted to a merger rate as soon as one knows how long a merger 
is identifiable as either a galaxy pair or morphologically disturbed galaxy. Hence, both methods need a timescale for which mergers are identified as such. 
In this series of papers, we focus on the morphological approach, specifically the morphology of galaxies in the 21 cm emission line (\hi).
Despite the generally lower resolution of the \hi \ maps compared to optical images, we argue that the morphological signature of a merger event can be equally or better observed because (a) the \hi \ disk extends well beyond the stellar disk (i.e., offers the same number of resolution elements as the optical disk), and (b) the atomic gas is disturbed well before the stellar disk. Therefore, \hi \ morphology looks promising as an alternate tracer of mergers, provided representative enough volumes are surveyed.

A popular measure of the merger signal in galaxy morphology uses the Concentration-Asymmetry-Smoothness parameters \citep[CAS,][]{CAS}, but additional parameters such as Gini (G) and $M_{20}$ \citep{Lotz04} are used extensively as well. The merger signal is usually found in the restframe ultraviolet or B-band, calibrated by galaxies in the local Universe. The timescale on which a merger is identifiable is found from CAS or G and $M_{20}$ measurements of N-body simulations of mergers \citep{Conselice06b,Lotz08a,Lotz10a,Lotz10b} or alternatively from the decline in identified mergers between two nearby redshift bins, assuming only passive evolution in merger rate \citep{Conselice09c}. Both approaches give consistent estimates of the merger time-scale of about a Gyr or less, allowing for estimates of the typical number of mergers a massive galaxy undergoes during its lifetime \citep{Conselice06b, Conselice09c, Lotz08a}. 
A limitation is that morphological identification of mergers is most effective for gas-rich galaxies \citep{Lotz10a} and varies with the mass ratio of the merging galaxies \citep[e.g.,][]{Lotz10b}. 

All previous work on quantified galaxy morphology as a merger tracer ha focused on rest-frame ultraviolet or the blue side of optical. The benefits are clear for this technique as the higher redshift galaxies can be observed with the Hubble Space Telescope at these wavelengths at comparable spatial resolution to the local reference samples observed by GALEX or the Sloan Digital Sky Survey (SDSS). Star formation triggered by the merger increases surface brightness, easing their identification at higher redshift. 

However, in the coming decade, a new window on gas-rich mergers will be opening up with the commissioning of new radio telescopes and instruments: the South African Karoo Array Telescope \citep[MeerKAT;][]{meerkat1, MeerKAT,meerkat2}, the Australian SKA Pathfinder \citep[ASKAP;][]{askap2, askap1, ASKAP, askap3,askap4}, the Extended Very Large Array \citep[EVLA;][]{evla} and the APERture Tile In Focus instrument \citep[APERTIF;][]{apertif,apertif2} on the Westerbork Synthesis Radio Telescope (WSRT). Ultimately, this investment in radio observatories will culminate in the Square Kilometer Array \citep[SKA;][]{ska}. These observatories will extend the 21 cm line observations of galaxies to high redshift and low column densities. Specifically, the effective all-sky survey of the nearby Universe with ASKAP (the WALLABY project, Koribalski et al. {\em in preparation}) and the Medium Deep Survey with WSRT/APERTIF spans an ideal volume for a local merger fraction and rate measurement. 
The \hi \ disk's morphology can be an alternative tracer of the merger fraction and consequently merger rate. There is already ample anecdotal evidence for the sensitivity of \hi \ morphology to interaction \citep[e.g., ``The \hi \ Rogue Galaxy Gallery",][\url{http://www.nrao.edu/astrores/HIrogues/}]{Hibbard01}, and more quantified relations between density and \hi \ morphology exist \citep[e.g.,][]{Bouchard09}.

In this series of papers, we explore how well the parameterized \hi \ morphology trace mergers in two local \hi \ surveys: the \hi \ Nearby Galaxy Survey \citep[THINGS,][]{Walter08} and the Westerbork \hi \ Spiral Project \citep[WHISP,][]{whisp1,whisp2}. We have found that the \hi \ column density maps -- while at typically lower spatial resolution than the optical wavelengths-- are just as sensitive if not more so to the effects of a merger, especially when expressed in the morphological qualifiers customarily used in optical or UV classification \citep{Holwerdapra09, Holwerda10b}. 
For the first time, it may be possible to select ongoing merging galaxies from their \hi \ morphology without any need for a human observer. This will greatly simplify the classification of galaxies and mergers in the upcoming all-sky \hi \ surveys. 

In a subset of the WHISP sample we showed that there is a simple cut in morphological parameter space that divides the merging disks from the general non-merging population \citep{Holwerda10c}. The question remains, however, how long a merger is identifiable in this new window before we can convert an observed merger fractions in WHISP to a merger rate \citep[our fifth paper,][]{Holwerda10e}. In this paper, we explore the merger visibility timescale using N-body simulations of $L^*$ disks with cold gas disks, both evolving passively and undergoing a violent 1:1 merger with a similar disk.
The organisation of this paper is as follows: section \ref{s:morph} discusses the morphological parameters we use briefly, section \ref{s:sims} describes the simulations we used, how the \hi \ map was generated, and the limitations of these simulations and the \hi \ perspective. Section \ref{s:results} is a discussion of our results from the N-body simulation, and section \ref{s:concl} our conclusions.

\section{Morphology}
\label{s:morph}

In this series of papers we use the CAS system (Concentration-Asymmetry-Smoothness) from \cite{Bershady00,Conselice00a,CAS}, the Gini/$M_{20}$ system from \cite{Lotz04} and our own parameter $G_M$, the Gini parameter of the second order moment \citep[][]{Holwerda10c}.
Concentration is defined as:
\begin{equation}
C = 5 ~ log \left( {r_{80} \over  r_{20} } \right)
\label{eq:c}
\end{equation}
\noindent where $r_{\%}$ is the radius which includes that percentage of the intensity of the object.
In an image with $n$ pixels with intensities $I(i,j)$ at pixel position $(i,j)$ and $I_{180}(i,j)$ is the value of the pixel in the rotated image, Asymmetry is:
\begin{equation}
A = {\Sigma_{i,j} | I(i,j) - I_{180}(i,j) |  \over \Sigma_{i,j} | I(i,j) |  },
\label{eq:a}
\end{equation}
\noindent and Smoothness is defined as:
\begin{equation}
S = {\Sigma_{i,j} | I(i,j) - I_{S}(i,j) | \over \Sigma_{i,j} | I(i,j) | },
\label{eq:s}
\end{equation}
\noindent where $I_{S}(i,j)$ is the same pixel in the image after smoothing. 

The Gini parameter is an economic indicator of equality, i.e., G=1; all the flux in one pixel, G=0; equal values for all pixels in the object. We use the implementation from \cite{Lotz04}:
\begin{equation}
G = {1\over \bar{I} n (n-1)} \Sigma_i (2i - n - 1) | I_i |,
\label{eq:g}
\end{equation}
\noindent where $I_i$ is the intensity of pixel $i$ in an flux-ordered list of the $n$ pixels in the object, and $\bar{I}$ the mean pixel intensity.

They also introduced the $M_{20}$ parameter:
\begin{equation}
M_{20} = \log \left( {\Sigma_i^k M_i  \over  M_{tot}}\right), ~ {\rm for ~ which} ~ \Sigma_i^k I_i < 0.2 ~ I_{tot} {\rm ~ is ~ true}.\\
\label{eq:m20}
\end{equation}
\noindent where $M_{tot} = \Sigma M_i = \Sigma I_i [(x_i - x_c)^2 + (y_i - y_c)^2]$. The center of the object is at ($x_c$,$y_c$).
and pixel $k$ is the pixel marking the top 20\% point in the flux-ordered pixel-list.

Instead of the intensity of the pixel ($I_i$) one can use the second order moment of the pixel ($M_i =  I_i [(x_i - x_c)^2 + (y_i - y_c)^2]$) in equation \ref{eq:g}. This is our $G_M$ parameter:
\begin{equation}
G_M = {1\over \bar{M_i} n (n-1)} \Sigma_i (2i - n - 1) | M_i |,
\label{eq:gm}
\end{equation}
\noindent which is an indication of the spread of pixel values with distance to the galaxy centre.  
%
%
The combination of these parameters quantify the morphology of \hi \ column density maps. We refer the reader for a broader discussion of these parameters in the previous papers of this series \citep{Holwerda10b, Holwerda10c}. 

As input these parameters need an estimate of the centre of the object and a definition of the area over which they need to be computed. The centres of the objects are a given, in the case of the simulations as the centre of the dark matter halo. In the case of surveys as the centre of the light distribution. 
To define the extent of the \hi \ disk , we use a threshold of $N_{HI} = 3 \times 10^{20}$ cm$^{-2}$. This defines the area over which the morphological parameters are computed. This threshold is the stated observational limit of the WHISP survey \citep{whisp1,whisp2}, and it is in between the two \hi \ column density thresholds we used in \cite{Holwerdapra09,Holwerda10b} for the extent of the the stellar and gas disks in the THINGS survey \citep{things}.

Starting with the snapshots of the \hi \ in simulations, we explore in this paper the evolution of the above morphological parameters over time, both during a violent merger as well as the passive evolution in a few isolated disks. 
We determine how long a merging disk spends in the morphological parameter space that hosts mergers as a function of viewing angle and physical disk characteristics. We found this space both from literature definitions for optical morphology and empirically in the WHISP sample for \hi \ morphology \citep{Holwerda10c}.
Visibility time scales can then be used to convert an observed fraction of merging galaxies into a merger rate --the number of galaxies merging in a given volume per Gyr-- as applied to the whole WHISP sample in the companion paper \cite{Holwerda10e}. One caveat that we need to point out from the outset, is that our merger simulations are for 1:1 merger of massive spiral disks and the WHISP sample consists of a mix of galaxy masses ({\em see Appendix B
in the online version of the paper}). Future suites of merger simulations should include a wider mass range and gas fraction.

\section{Simulations of Merging and Isolated Spiral Disks}
\label{s:sims}

Our primary set of simulations is a suite of equal mass mergers from \cite{Cox06a}, from which an 
atomic hydrogen component was estimated by one of us (T.J. Cox) assuming thermal equilibrium.
\hi \ maps for a control sample of isolated spiral disks, passively evolving for 2.5 Gyr, were constructed 
in the same manner.
 
To gauge the importance of different implementations of a Milky Way size disk in a merger simulation, we also 
use a single merger simulation and isolated disk simulation  from \cite{Weniger09}, also converted to an \hi \ column density map.
The code, assumed physics and timescales are different for the Weniger simulations (see \S \ref{ss:weniger}).

All the merger simulations are for equal mass mergers of two large spiral galaxies. 
The suite from Cox et al. and the single simulation from Weniger et al. are for spiral galaxies of similar scale and mass (see Tables \ref{t:mass} and \ref{t:scales}).

\subsection{Simulations from Cox et al. 2006}
\label{ss:cox}

The merger simulations suite from \cite{Cox06a} use the N-body/SPH code GADGET \citep{Springel01, Springel02}. 
It conserves entropy and features additional routines that track the 
radiative cooling of gas and star formation (Volker Springel, private communication to T.J. Cox in 2001). 
In addition, \cite{Cox06a} implemented several new features themselves: stellar feedback, metallicity-dependent 
cooling and the ability of each gas particle to spawn multiple new stellar particles.

The simulations are for the isolated disks as well as the major mergers with variations in disk properties and 
viewing angles (Table \ref{t:sims}). All these simulations run for 2.5 Gyr in steps of 100 Myr. 
Images are surface brightness values in  $10^4 ~ h ~ M_\odot ~ pc^{-2}$ with a pixelscale of 150 pc. 
The high resolution of these simulations enabled us to construct \hi \ maps with sampling much finer than any 
current survey (See Table B.1 ({\em in the online appendix}), 
and closer to the resolution typical for optical images.
The nominal disks in these simulations are spirals with a $\sim10^{9} M_\odot$ gas disk (unless noted differently), 
which, in the case of mergers, interacts with an identical disk.

Significant disk properties that are varied are mass ({\it high-m} and {\it low-m}), 
and the treatment of the feedback from star-formation on the interstellar matter ({\it ism2}). 
We consider three orientations: face-on, at a 45$^\circ$ angle ({\it orientation2}) and edge-on 
({\it edge-on}). Examples of the face-on disks, isolated and merger, are shown in Figure \ref{f:snap} and Appendix A ({\em online version}). For more details on these 
simulations, we refer the reader to \cite{Cox06a} and \cite{Cox06b}.

\begin{table*}
\caption{Mass fractions and number of particles in the different components in both simulations.}
\begin{center}
\begin{tabular}{l l l l l}
			& Cox et al 2006	&					& Weniger et al 2009	& \\
Component 	& Mass fraction		& Nr of particles		& Mass fraction			& Nr. of particles \\
\hline
\hline
Bulge-stars 	& 0.012			& 10 000		& 0.069 	& 10 000\\
Disk-stars 	& 0.048			& 30 000		& 0.118 	& 75 098\\
Clouds 		& 				& 			& 0.016 	& 36 347\\
Gas-particles 	& 0.065			& 30 000		& 0.008 	& 9017\\
DM-particles 	& 0.8	75			& 100 000		& 0.789 	& 100 000\\
\hline
Total Mass 	& 8.12 $\times 10^{11} M_\odot$ 			& 170 000 	& 2.46 $\times 10^{11} M_\odot$	& 230462 \\
\end{tabular}
\end{center}
\label{t:mass}
\end{table*}%


\begin{table}
\caption{Scales of the different components of the disk galaxies.}
\begin{center}
\begin{tabular}{l l l }
						& Cox et al 2006	& Weniger et al 2009 \\
Component 				& (kpc)			& (kpc) \\
\hline
\hline
Disk						&				& \\
 - scalelength, $R_d$ 		& 4.0				& 4.0 \\
 - scaleheight, $z_0$ 		&1.0 				& 0.4\\
Gas scalelength			& 16.5			& \\
Bulge scalelength, $R_b$ 	& 0.45 			&  \\
\hline

\end{tabular}
\end{center}
\label{t:scales}
\end{table}%

\begin{table}
\caption{Cox et al Suite of Merger Simulations.}
\begin{center}
\begin{tabular}{l l }
\hline
\hline
isolated 			& \\
\hline
face-on    			& reference, $3.34 \times 10^{10} M_\odot$ total disk mass \\
edge-on   		 	& same but edge-on perspective\\
high-mass  		& face-on perspective on $8.7 \times 10^{10} ~ M_\odot$  \\
low-mass   		& face-on perspective on $1.24 \times 10^{10} ~ M_\odot$  \\
\hline
\hline
 mergers			& \\
\hline
{\it face-on} 			& \\
\ run 1        		& face-on perspective on two \\
				& $3.34 \times 10^{10} ~ M_\odot$ disks collision \\
\ run 2        		& same with slightly different collision orbit.\\
{\it orientation}		& \\
\ 45$^\circ$		& same at a 45$^\circ$ camera angle \\
\ edge-on			& same for edge-on camera \\
{\it disk mass}			& \\
\ high-mass		& face-on perspective on two $8.7 \times 10^{10} M_\odot$  \\
\ low-mass		& face-on perspective on two $1.24 \times 10^{10} M_\odot$  \\
{\it different ISM}		& different ISM feedback treatment, \\
				& see \cite{Cox06a}\\
\ high-mass 		& with $8.7 \times 10^{10} M_\odot$ disks.\\
\ low-mass  		& with $1.24 \times 10^{10} M_\odot$ disks.\\
\hline

\end{tabular}
\end{center}
\label{t:sims}
\end{table}%

\subsection{The single merger simulation from Weniger et al (2009)}
\label{ss:weniger}

As a consistency check, we apply our morphological code to the single merger simulation from \cite{Weniger09}, 
as well as a single isolated disk simulation.
Following \cite{Harfst06}, they constructed the initial galaxies by firstly generating a disk/bulge/halo - system using
the method of \cite{Kuijken95}. In their next step, one fifth of the stellar disk particles were transformed
into molecular clouds and finally the diffuse ISM was added as an initially slowly rotating homogeneous sphere, which
will collapse in the first 200 Myr forming a warm disk. The collision-less model of \cite{Kuijken95} realises an 
equilibrium configuration but in this case, the equilibrium is affected by adding 
the ISM. Therefore, \cite{Weniger09} first follow the systemÕs evolution, until a quasi-equilibrium is established. 
The numerical integration is done by means of a TREE-SPH code combined with the sticky particle method 
\citep{Theis93}. Gravitational forces are determined by the DEHNEN-Tree \citep{Dehnen02}. 
The cold gas ($T < 10^4 K$) in this simulation was converted to the \hi \ map, assuming that the fraction of the 
cold gas which is observable as \hi, depends on the thermal pressure. The HI map was smoothed by a gaussian
with a FWHM of 1 kpc. The centre of the interacting disk is taken to be the centre of the bulge stars distribution.
Time steps are 10 and 50 Myr for the isolated disk and merger simulation, respectively. For more details we refer 
the reader to \cite{Harfst06} and \cite{Weniger09}.

\subsection{Limitations}
\label{ss:limitations}

The suite of simulations of isolated and merging disks form \cite{Cox06a} as well as the implementation check from \cite{Weniger09} are limited in their scope of disk (and merger) properties, i.e., only a large, gas-rich spiral disk colliding with its twin or evolving passively. There are only three different disk masses and camera angles, and no change in merger impact parameter of incident angle (see Table \ref{t:sims}). 
We limited our scope to this suite as we could easily obtain \hi \ maps for these simulations and still be consistent in treatment of physics and particle resolution across the sample (with the Weniger implementation as the exception).
Our second reason was that simulations of  isolated disks with identical physics and resolution evolving for a similar amount of time were available. 

Future simulations of the evolution of \hi \ morphology should include more minor mergers, a greater range in disk mass, a range in gas fraction, different impact parameters and incident angles of the mergers, as well as more camera angles. 
The suite of simulations presented in \cite{Lotz10a,Lotz10b} explore parameter space more, generating optical morphology measures. A similar effort will be needed in \hi \ before volume-limited all-sky surveys in \hi  \ are completed to complement the initial effort presented in this paper.

\subsection{Shaping the \hi \ Morpholgy}
\label{ss:limitations}

The morphology of the \hi \ disk is influenced by many different processes, only one of which is gravitational interaction with a neighbouring galaxy. The morphology is also determined by the level of star-formation (triggered or not by the interaction), which consumes gas, and the stellar winds and supernovae feedback that create the typical "effervescent" look of \hi \ maps. 
An additional effect from star-formation (or an active AGN) can be that much of the gas in a galaxy is heated or ionized, and thus no longer visible in the \hi \ map. 
Depending on the temperature and dust fraction of the ISM, much of the higher density parts of the spiral gas disk will be in the form of molecular hydrogen, also removing it from the \hi \ map. Ram-pressure effects experience during travel through the intergalactic medium of a cluster may also contribute significantly to the appearance of the \hi \ disk. 

The simulations treat ISM feedback from star-formation comprehensively but for now, we set aside the effects of an AGN switching on or the interaction happening in a dense cluster medium.
In a further paper in this series \citep{Holwerda10d}, we do explore the how much ram-pressure and \hi \ stripping of galaxies change these parameters using the VLA Imaging of Virgo spirals in Atomic gas \citep[VIVA,][]{Chung09}. We find that stripped \hi \ disks are easily identified with low values of Concentration. The effect of moderate ram-pressure on a sample of \hi \ disks will, however, be more difficult to quantify. 
The phenomena of \hi \ lopsidedness, which may be related to ram-pressure stripping seems to have only a moderate effect on these parameters \citep[see][]{Holwerda10c}.

Therefore, we compare morphological evolution of \hi \ disks with the same physics (feedback, ISM heating, etc.), in identical disks, and only change one critical aspect: in some simulations these disks merge with a similar disks, in others, they evolve passively.

\begin{figure*}
\centering
\includegraphics[width=0.4\textwidth]{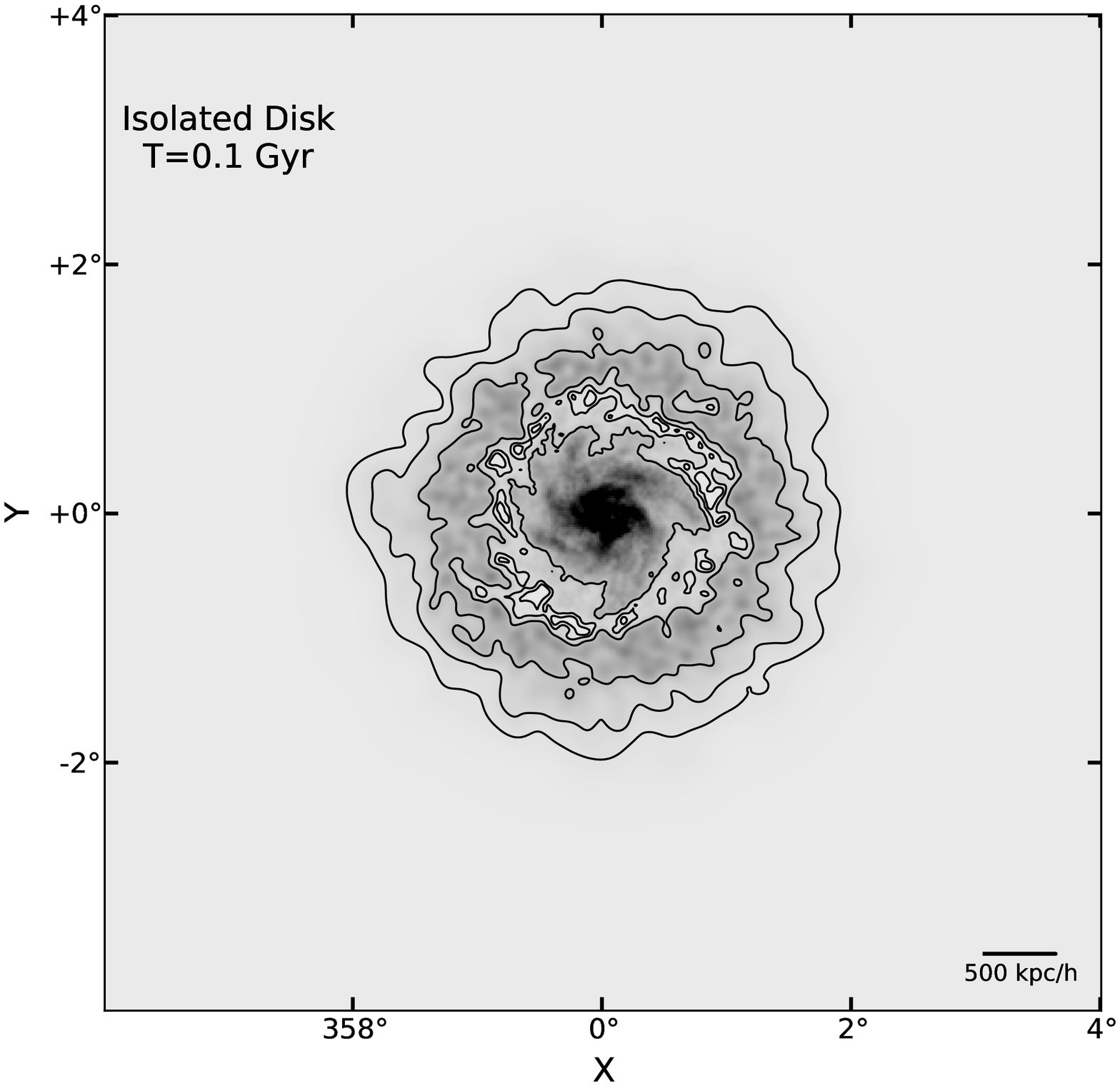}
\includegraphics[width=0.4\textwidth]{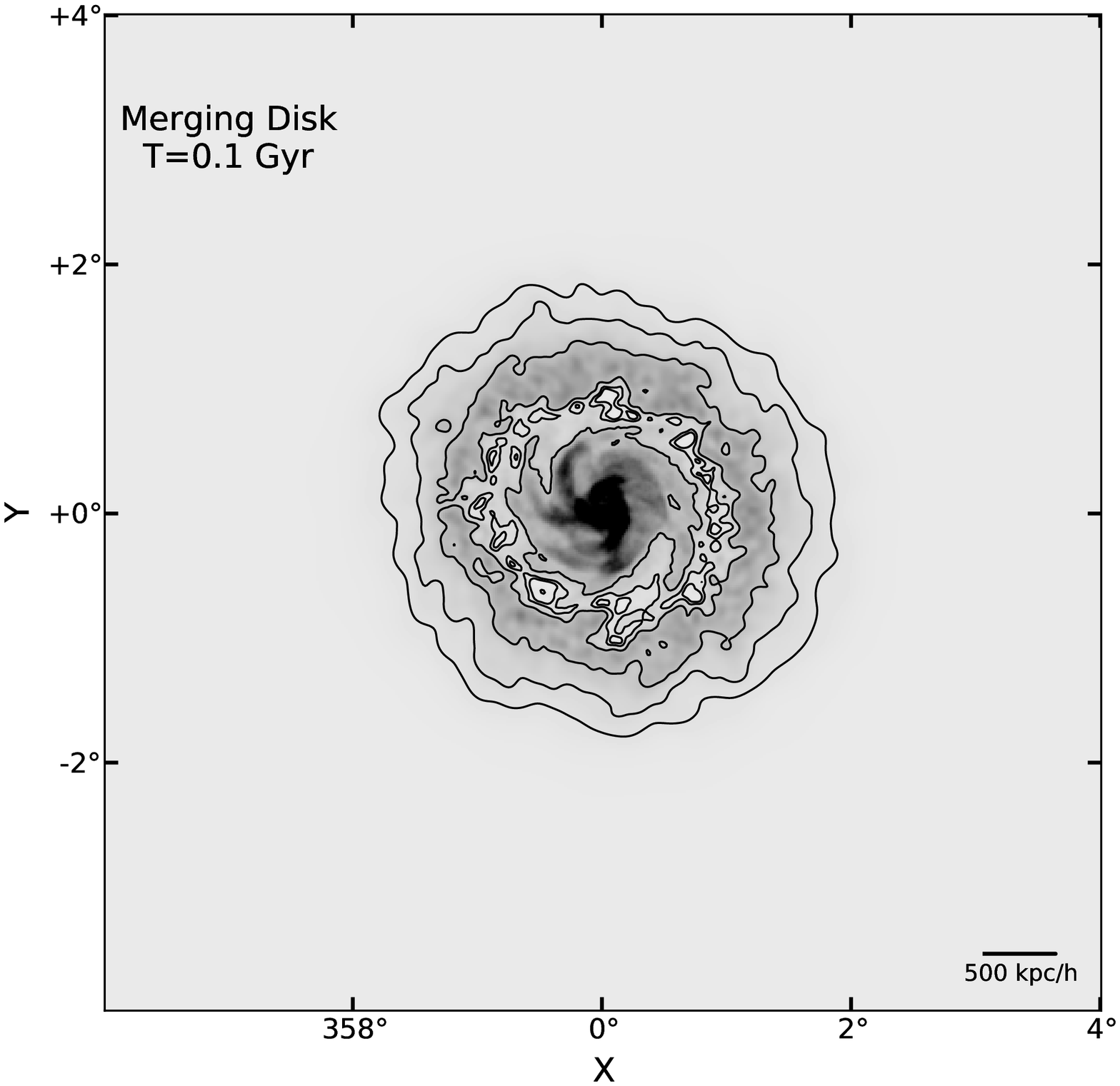}\\
\includegraphics[width=0.4\textwidth]{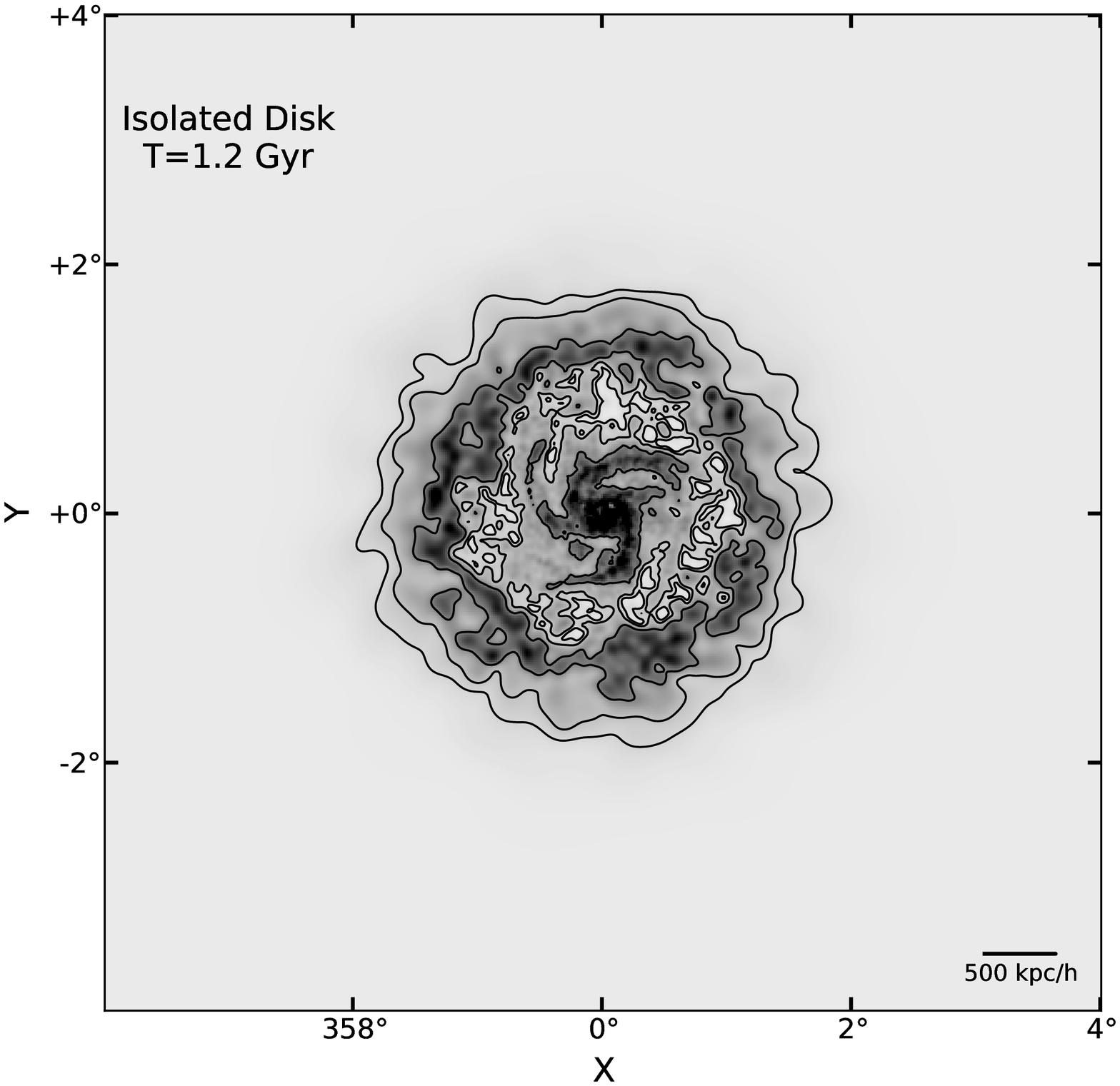}
\includegraphics[width=0.4\textwidth]{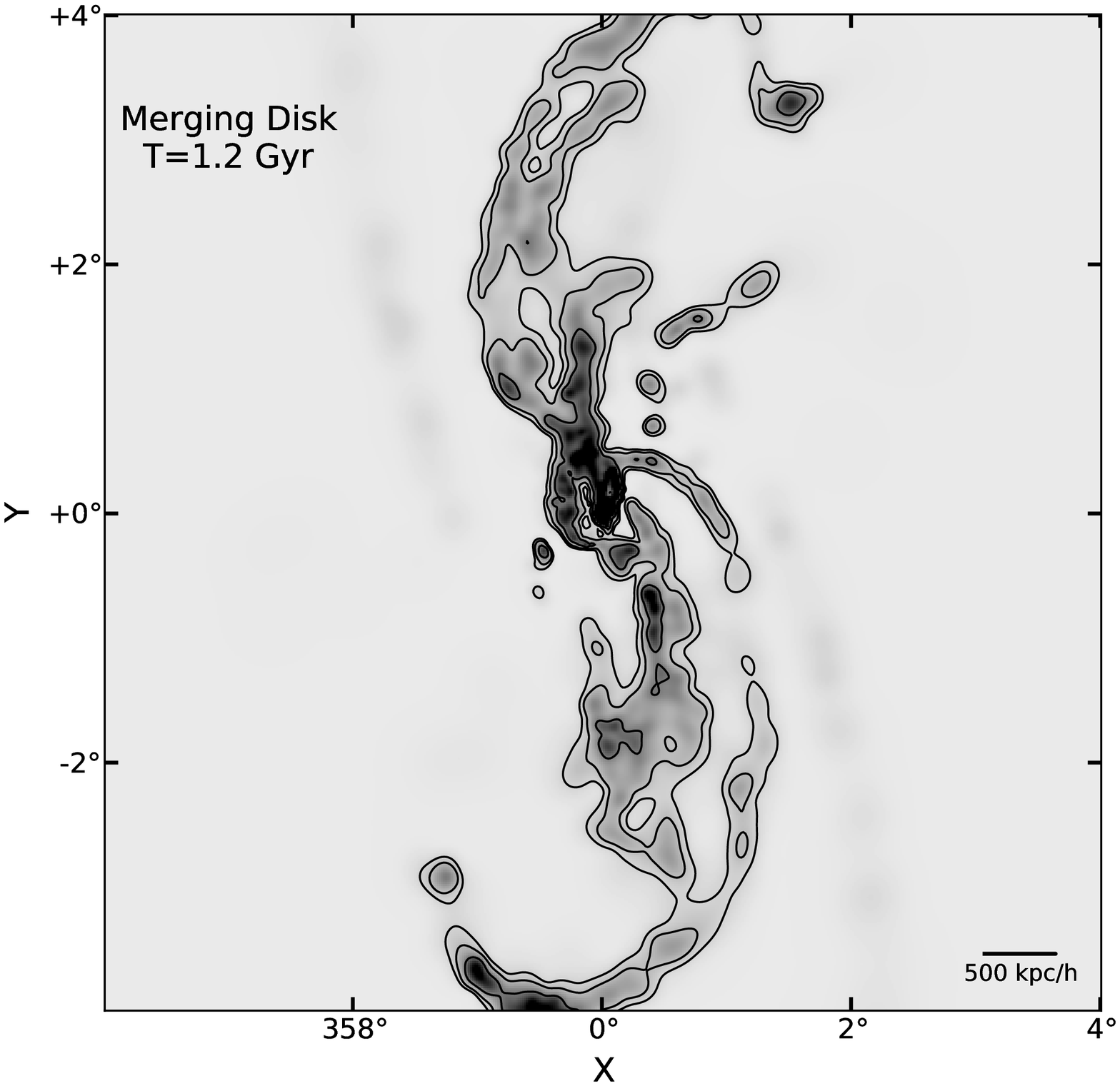}\\
\includegraphics[width=0.4\textwidth]{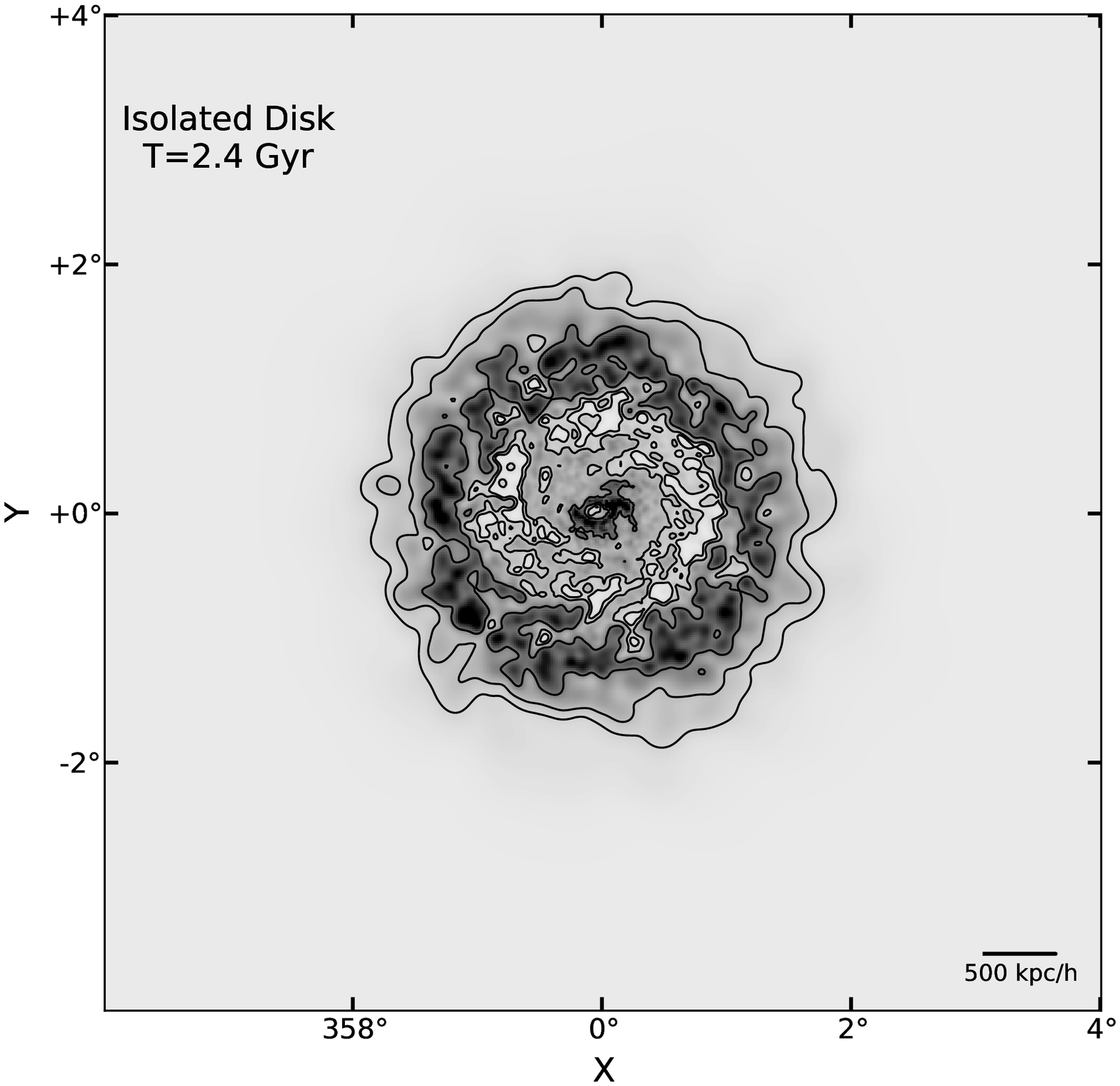}
\includegraphics[width=0.4\textwidth]{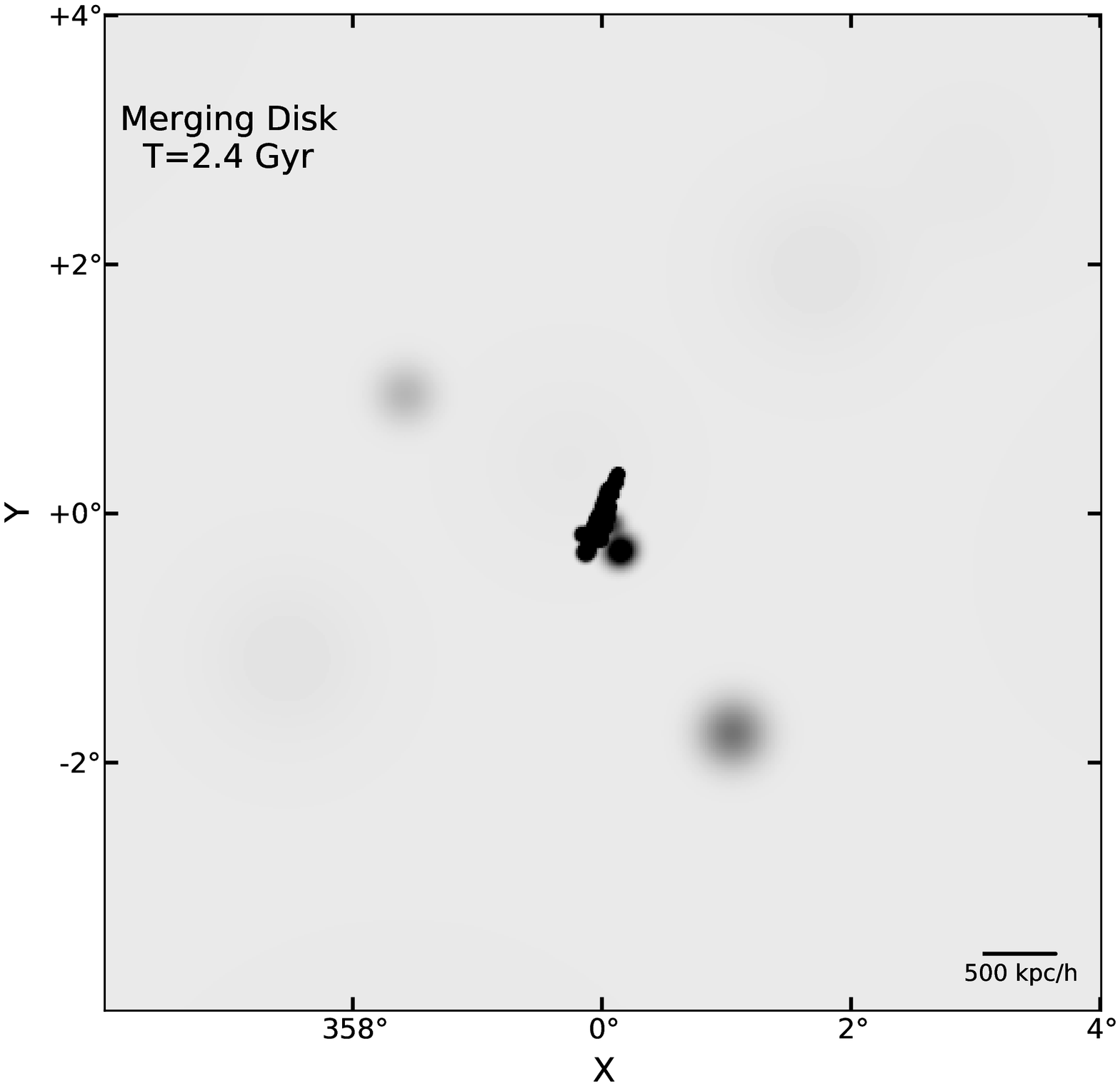}\\
\caption{\label{f:snap} Three \hi \ snapshots (at times T=0.1,1.2 and 2.4 Gyr) of two of our simulations, the nominal isolated, passively evolving spiral disk ({\bf left}) and the interacting and one instance of the merging disk ({\bf right}). Contours are at 0.5, 1.0, and 2.5 $M_\odot pc^{-2}$ to highlight the extent of the disks at low column densities.}
\end{figure*}

\section{\hi \  Morphology of Major Mergers}
\label{s:results}

We have run four isolated disks and eight merger simulations over a time of 2.5 Gyr \citep{Cox06a} and a single simulation of a major merger and one isolated face-on disk that ran for 3 Gyr \citep[][]{Weniger09}. 
The resulting tables of morphological parameters over time can be found in Appendix B ({\em online version}) for both isolated and merging disks.
The default perspective is face-on, with variation in disk mass and treatment of the ISM (Table \ref{t:sims}). 

\subsection{Merging and Isolated Disks}
\label{ss:iso}

\begin{figure*}
\centering
\includegraphics[width=0.32\textwidth]{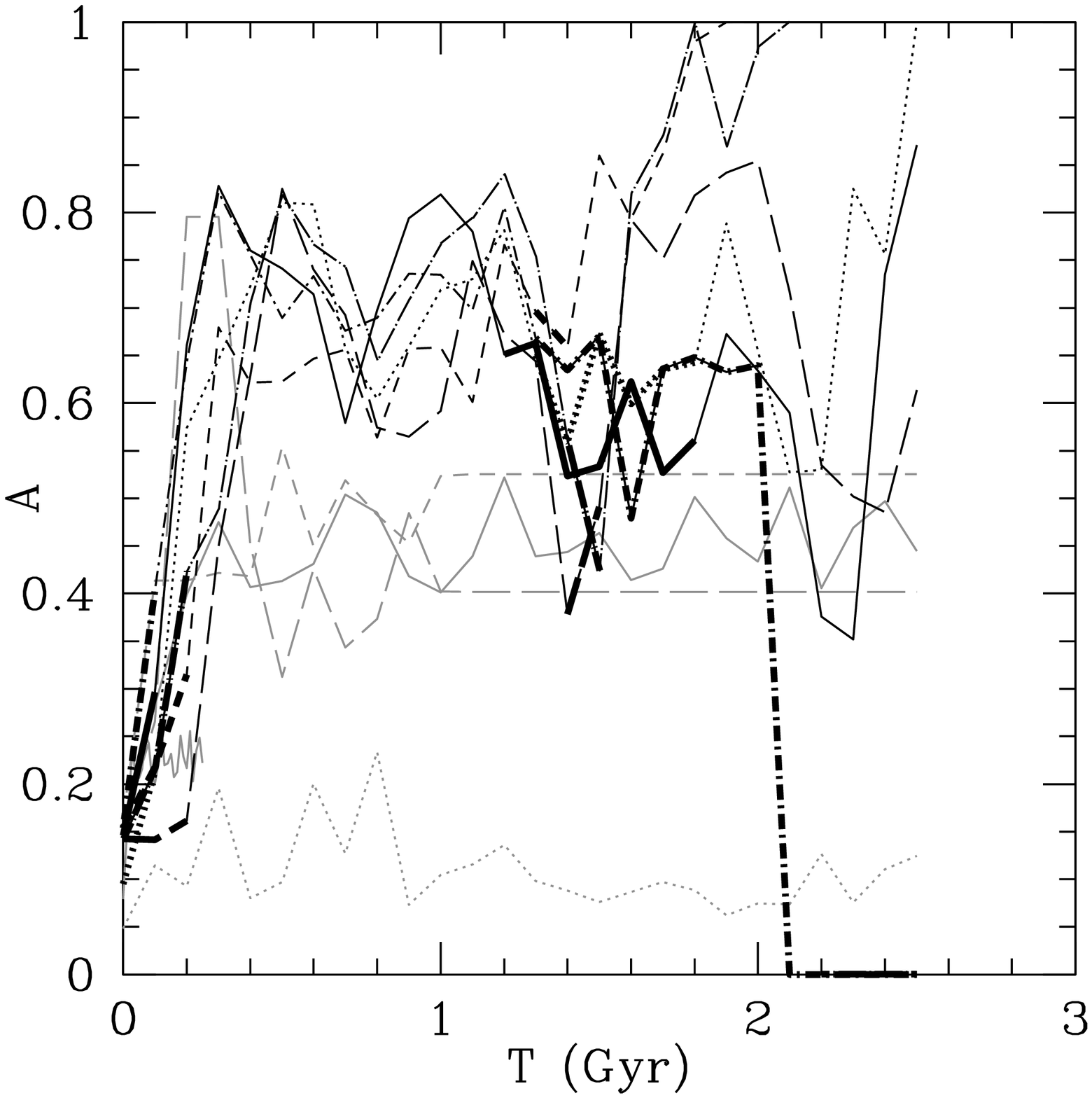}
\includegraphics[width=0.32\textwidth]{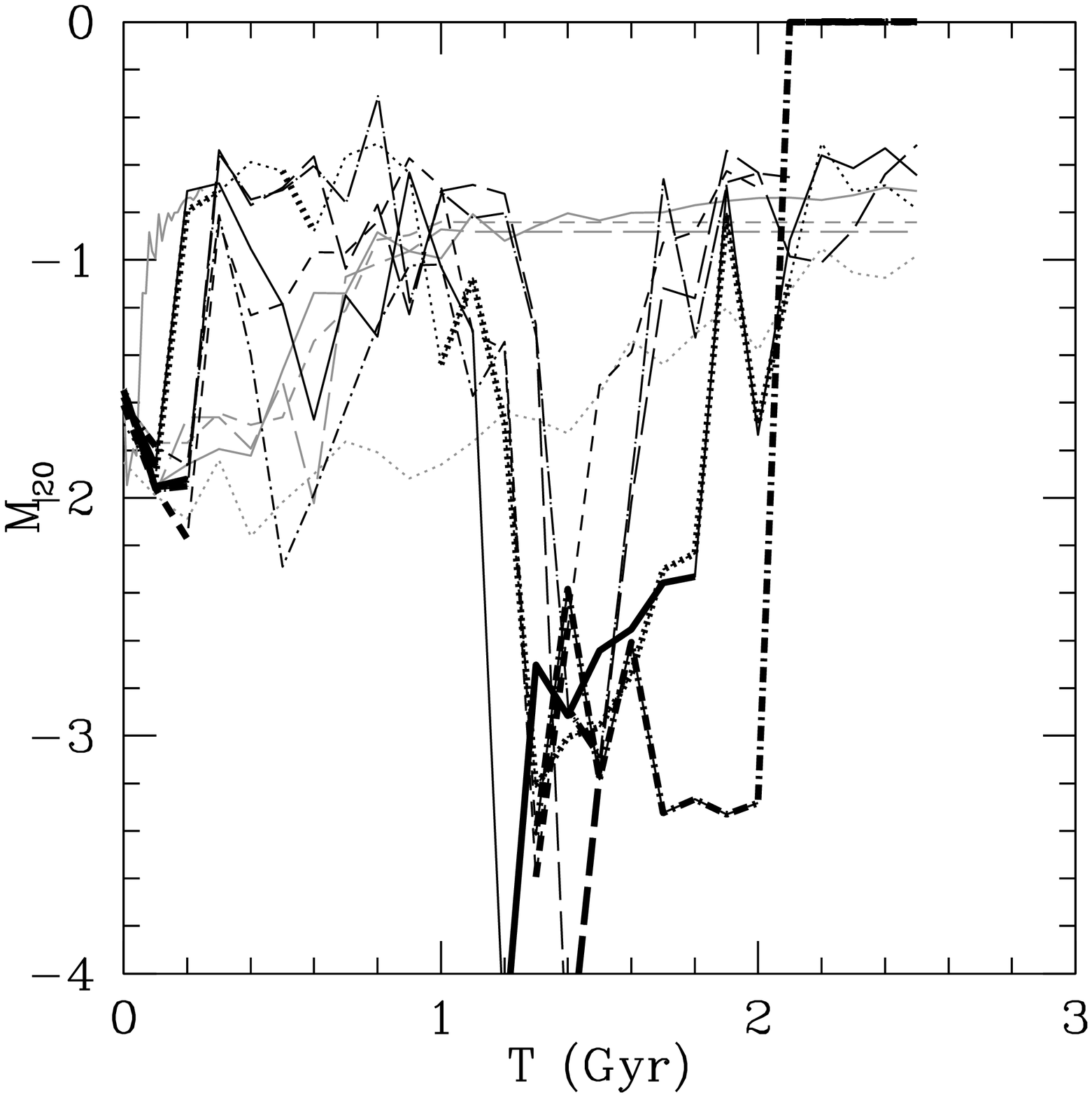}
\includegraphics[width=0.32\textwidth]{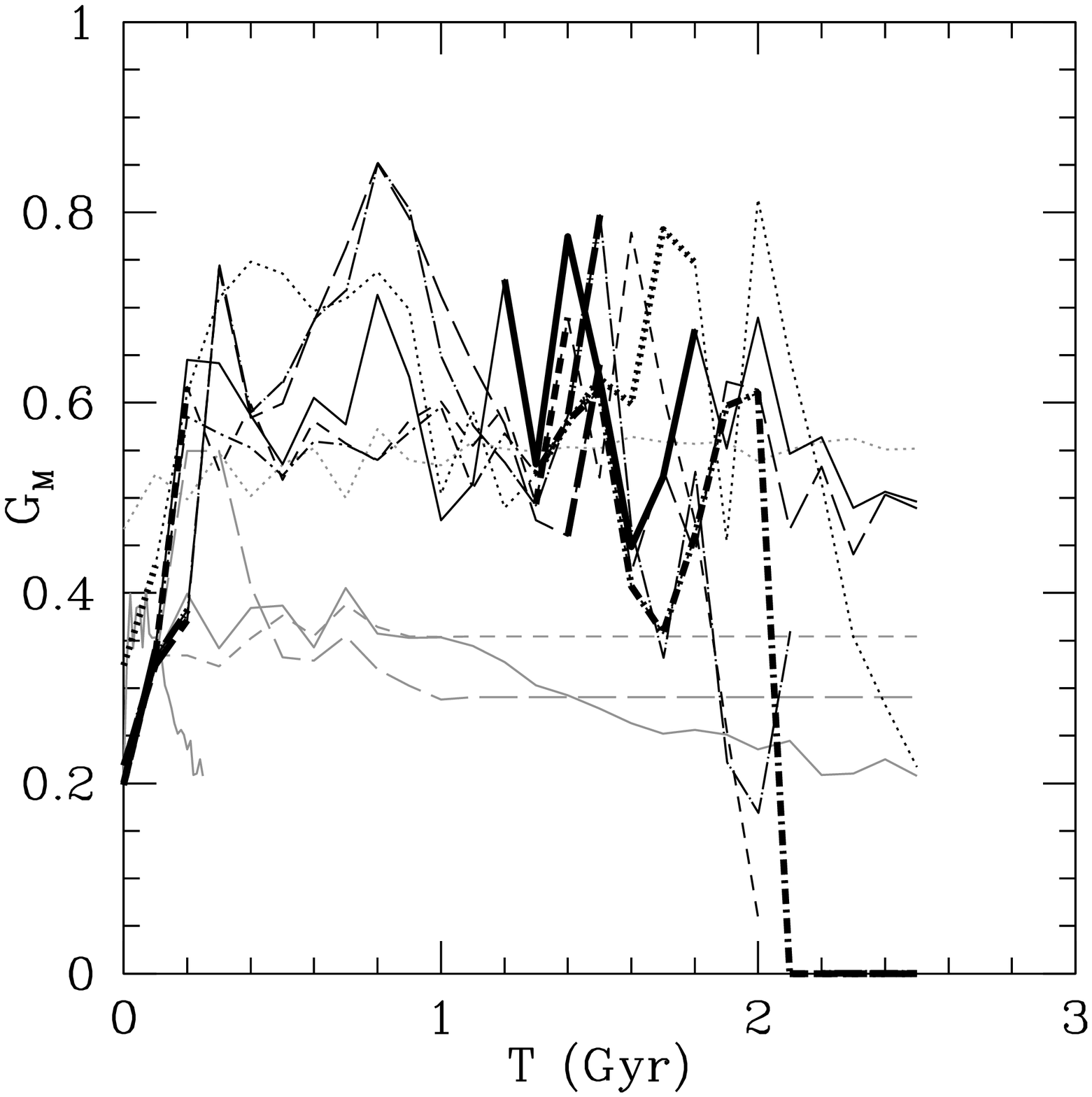}

\caption{\label{f:HI} Asymmetry, $M_{20}$ and $G_M$ as a function of time in the isolated (gray) and mergers (black) for the \hi \ disks from \protect\cite{Cox06a,Cox06b}. The disks are differentiated between nominal disks (solid lines), higher mass (short dashed lines), lower mass (long dashed lines), with different ISM treatment (high-mass; dot-short dash, low-mass; dot-long dash), and edge-on (dotted lines). The thick lines are the times $A < -0.2 \times M_{20} +  0.25$, one of our selection criteria. Isolated disks (gray lines) have less Asymmetry ($A<0.5$), high values for $M_{20}$ (close to 0) and a low $G_M$ values. Based on these timelines, the mergers do occasionally return to quiescent parameter space but differentiate well for a large fraction of the time. }
\end{figure*}

Figure \ref{f:HI} shows the progression over time of three parameters for both the merger scenarios (black lines) as well as the isolated disks (gray lines) for the \hi \ disks. 
We note how these three parameters already separate isolated disks and the merging ones as well as identify the times the galaxies are interacting the strongest.

All three parameters show a clear parameter space where the galaxy disk is almost certainly merging ($G_M > 0.6$, $M_{20} \sim -3$, and A $>$ 0.5) and a part where it is unlikely to be merging. In the case of $M_{20}$ and $G_M$, the merging disks return to the isolated disk's value periodically. Their values spike at the time of close passage of the merging companion. Variation of the disk mass ({\it low-m} and {\it high-m} simulations) do not change the overall morphological signature appreciably.

\subsection{Edge-on Disks}
\label{ss:edgeon}

\begin{figure*}
\centering
\includegraphics[width=0.32\textwidth]{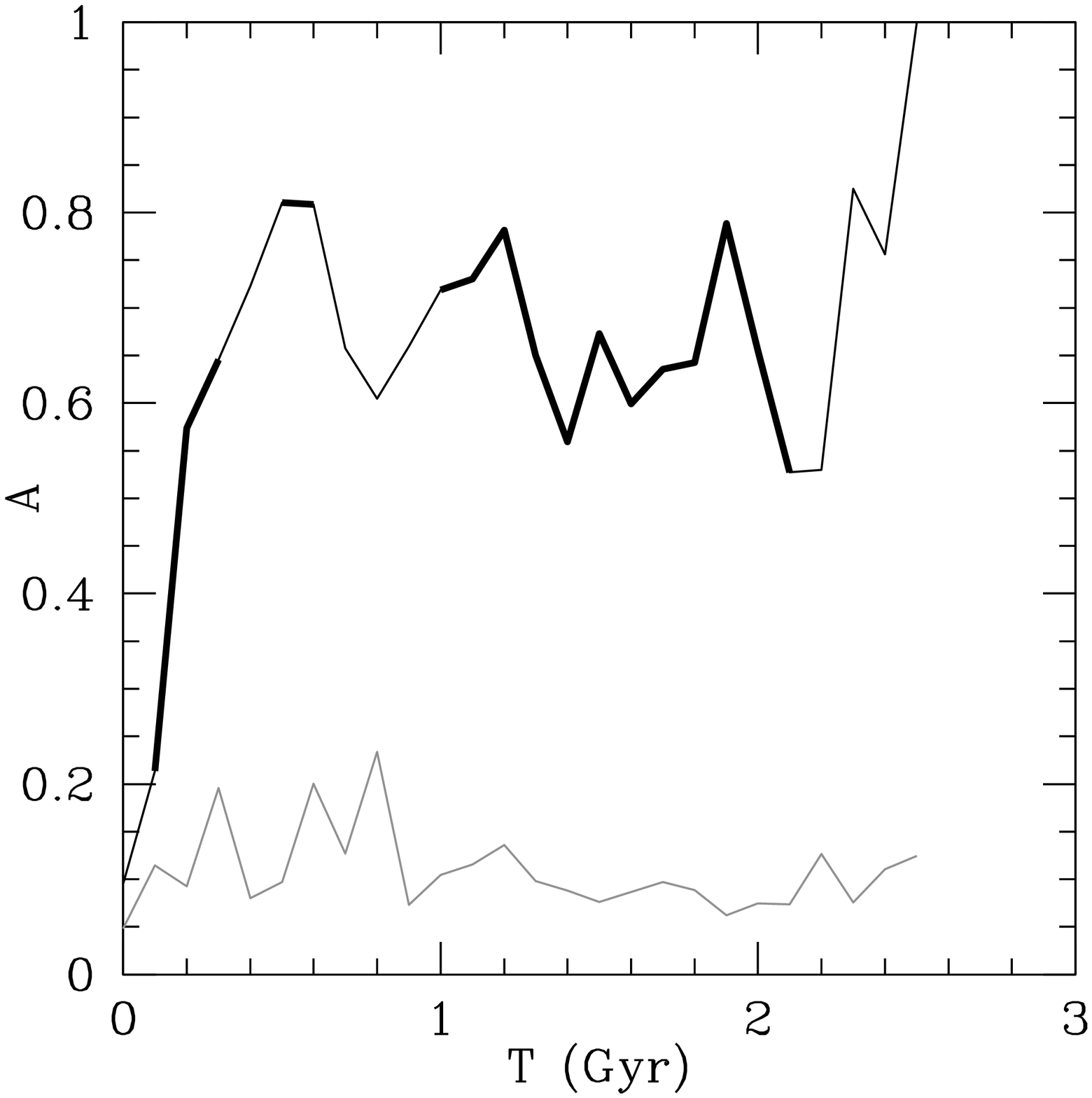}
\includegraphics[width=0.32\textwidth]{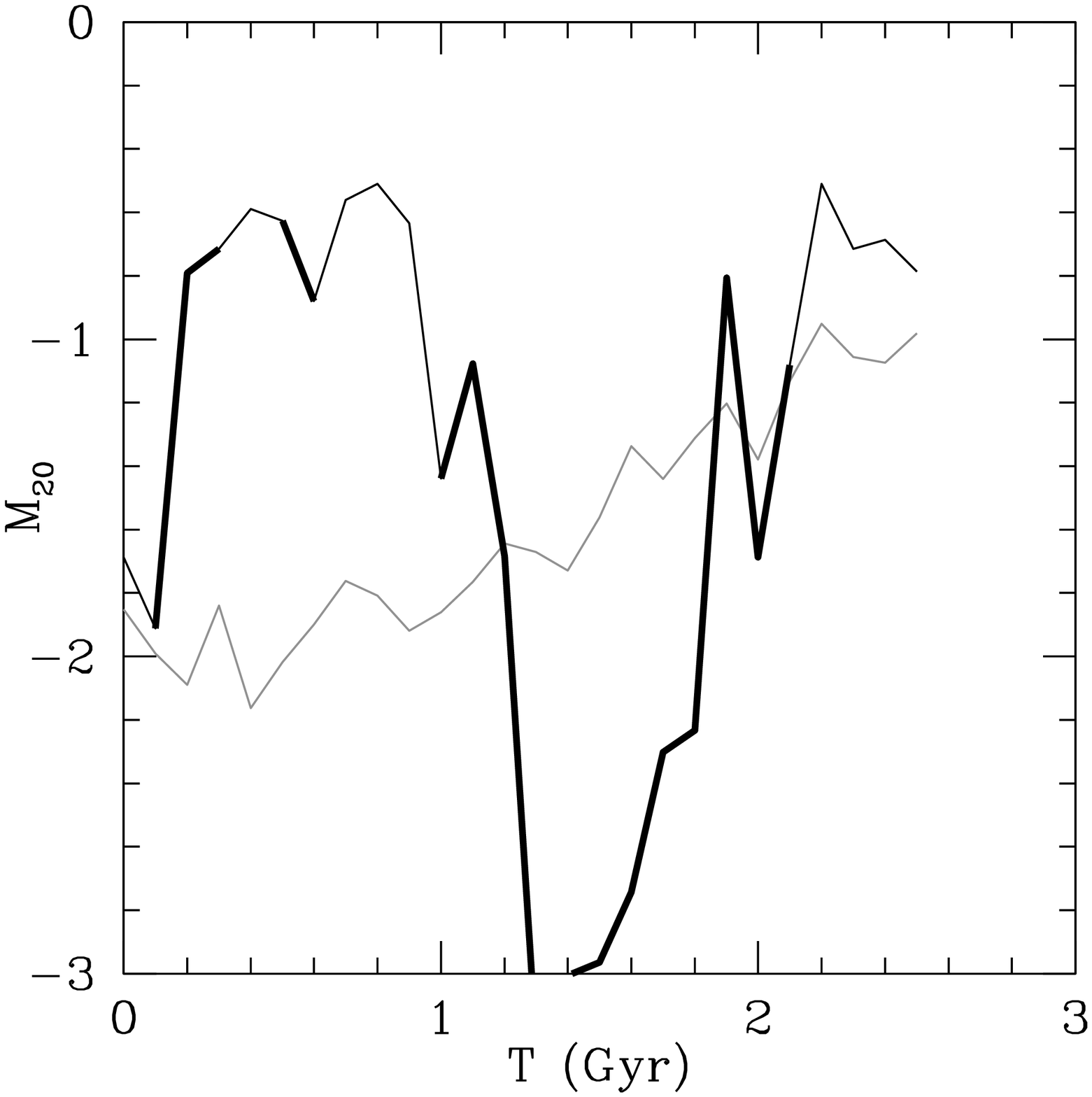}
\includegraphics[width=0.32\textwidth]{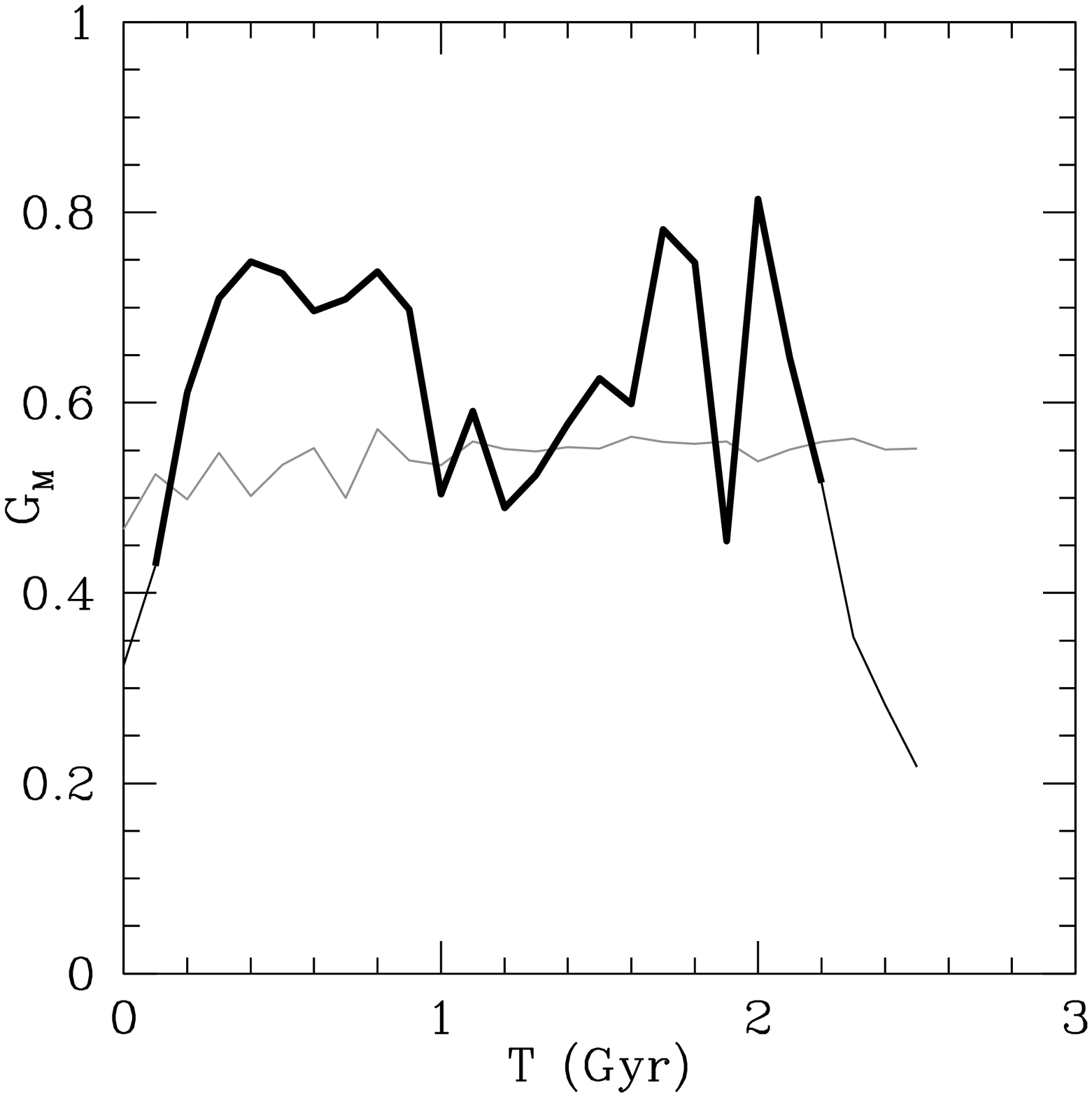}
\caption{\label{f:edgeon} {\bf Edge-on:} Asymmetry, $M_{20}$ and $G_M$ of the edge-on cases as a function of time in the isolated (gray) and mergers (black) of the cold gas. The thick lines are the times $A < -0.2 \times M_{20} +  0.25$, one of our selection criteria. }
\end{figure*}

Edge-on galaxies are morphologically the least disturbed perspective on a merger. Figure \ref{f:edgeon}~ shows the isolated disk and merging disk in cold gas, just from the edge-on perspective (dotted lines in Figure \ref{f:HI}). The merger is still morphologically selected by our selection criterion (see \S \ref{ss:mergcrit}), albeit not very often. However, in the case of the edge-on perspective, any signature in the {\em dynamical} profile of the galaxy will be the strongest, i.e., there will be a clear deviation of the typical ``double-horned" profile of the \hi \ line. Thus, our morphological approach and a kinematic approach to selecting interacting spirals in large \hi \ survey are very complementary. 

\subsection{Comparison to Weniger et al}
\label{ss:weniger}
\begin{figure*}
\centering
\includegraphics[width=0.32\textwidth]{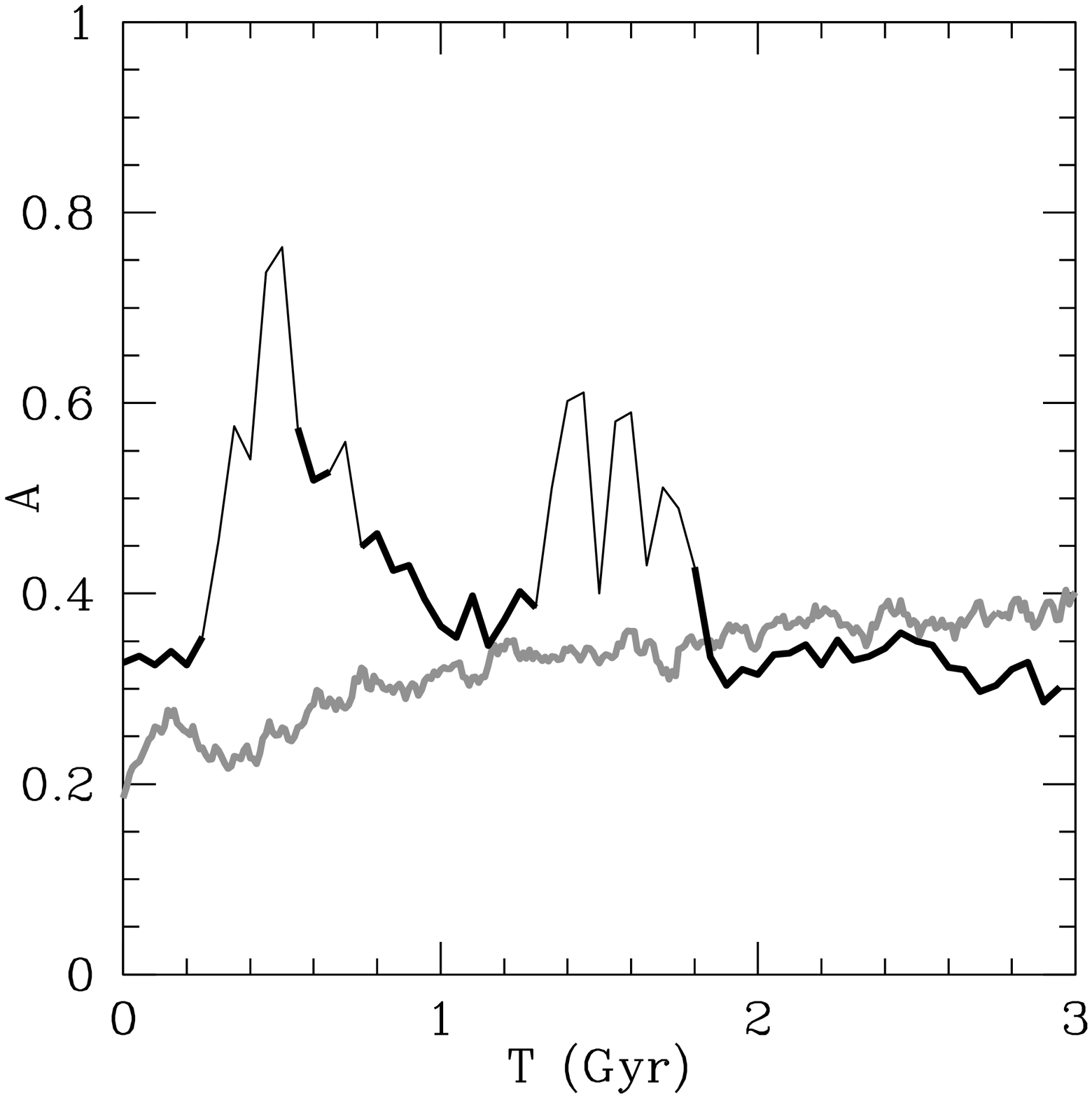}
\includegraphics[width=0.32\textwidth]{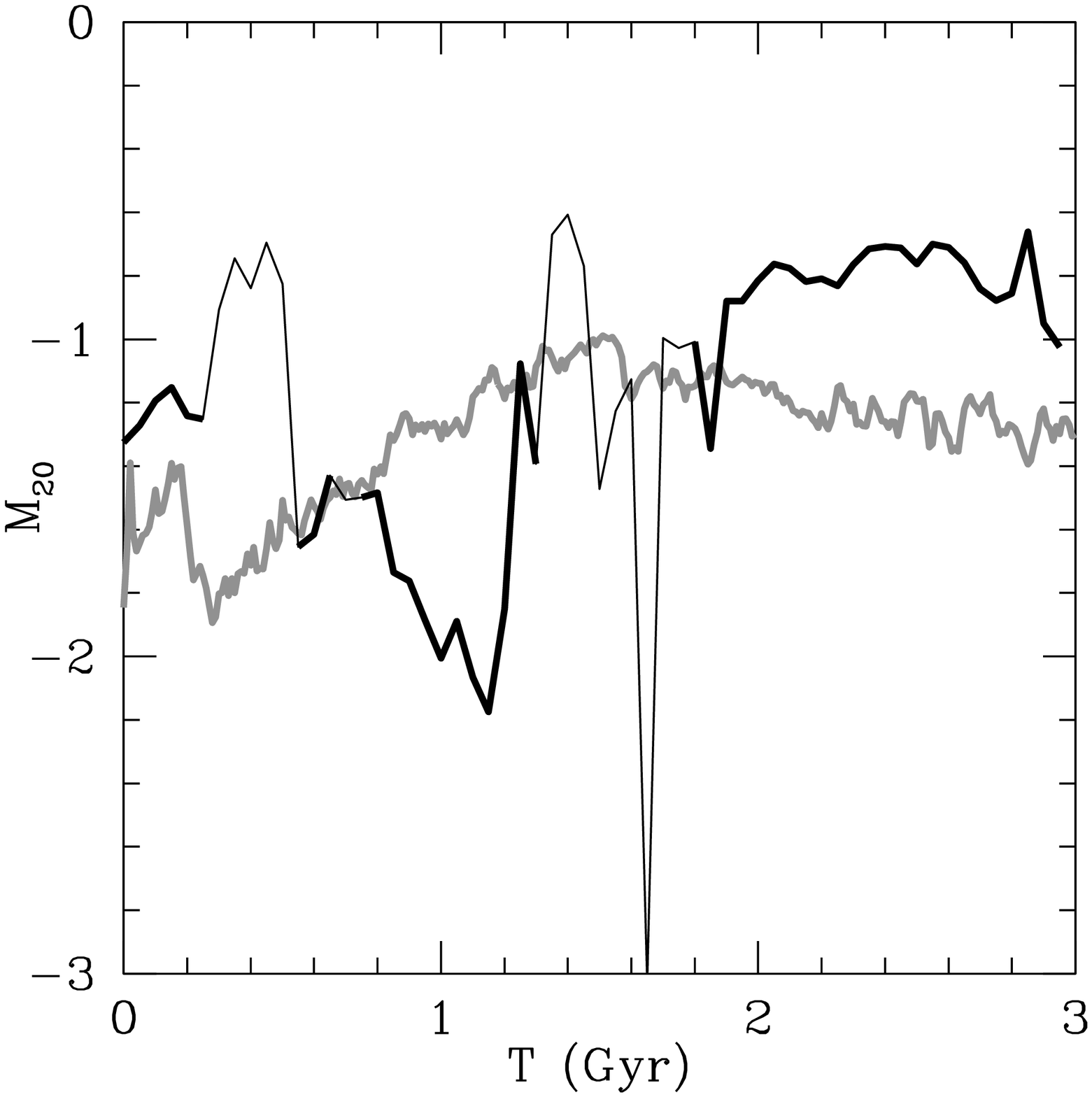}
\includegraphics[width=0.32\textwidth]{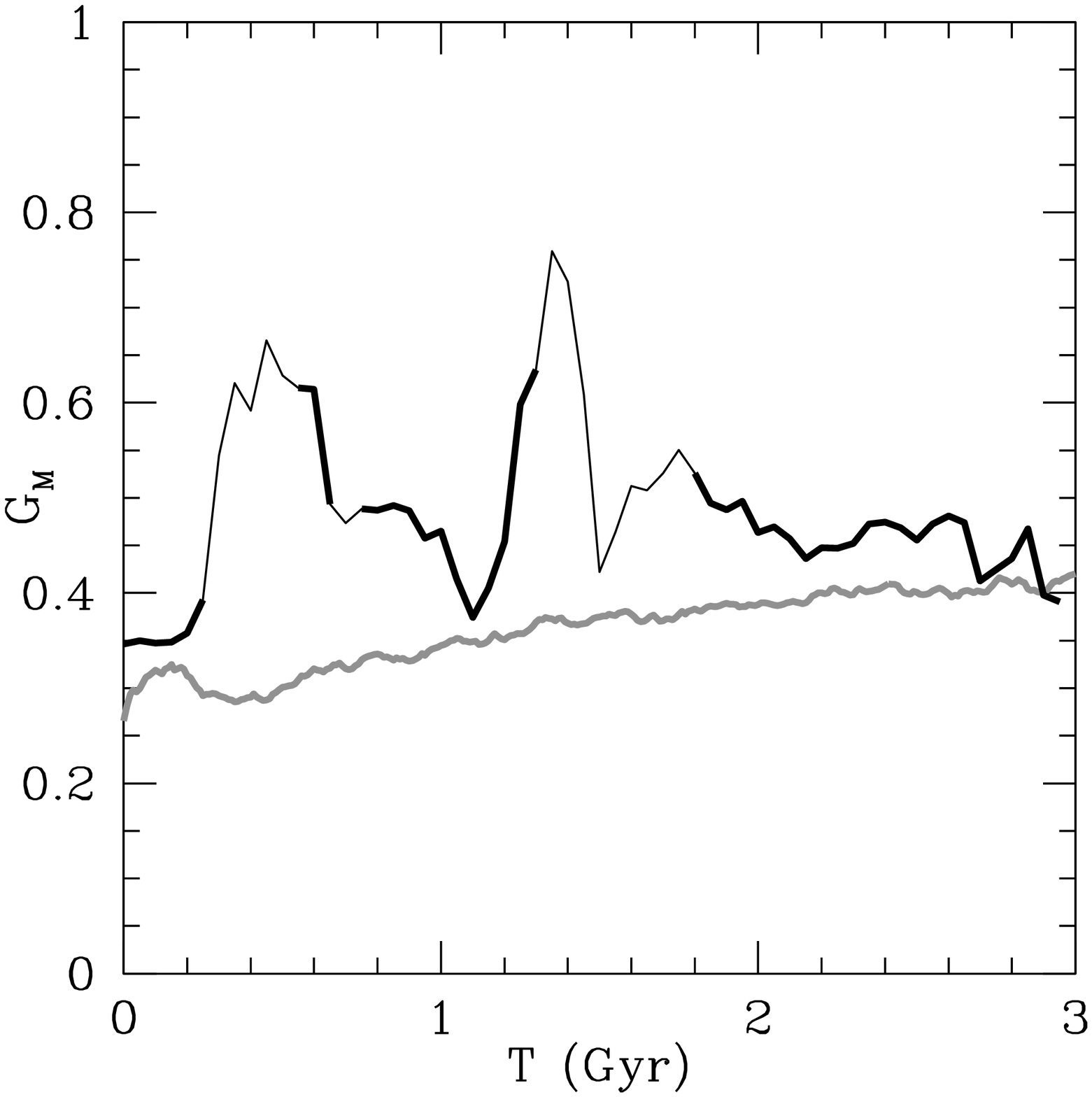}
\caption{\label{f:weniger} Asymmetry, $M_{20}$ and $G_M$ of both galaxies in the simulation of \protect\cite{Weniger09}. The morphology of the merging galaxy 
in \hi \ (black lines) and isolated galaxy (gray lines) are shown as a function of time. As an example, the thick black lines are those times of the simulation the merger 
would be selected by the $A < -0.2 \times M_{20} + 0.25$ criterion.}
\end{figure*}

Figure \ref{f:weniger} shows the values of Asymmetry, $M_{20}$ and $G_M$ of the \hi\ maps for a merging and an isolated galaxy from the simulation from \cite{Weniger09}. Asymmetry shows a steady rise with time for the isolated disk and clear spikes during the major encounters. $M_{20}$ displays a very similar evolution as the plot in Figure \ref{f:HI} with some notable evolution by the isolated disk. $G_M$ spikes at the close encounters, very similar to Asymmetry. 

The morphological selection criteria (see next section), perform well on the Weniger et al. simulation, selecting the merging galaxy for part of the time and mostly not selecting the isolated galaxy's \hi \ disk. The times, in billions of years, the Weniger et al merger are selected by the various morphological criteria are listed at the bottom of Table \ref{t:times}.
The general similar behavior of the Weniger et al. simulation to the suite from Cox et al. is an indication of the robustness of the morphological selection.

\subsection{Merger Selection Criteria}
\label{ss:mergcrit}

The goal of this paper is to obtain a typical time scale that mergers are visible as disturbed \hi \ morphologies. The visibility time depends on the selection criteria, 
observed wavelength and orientation of the merger. We can now explore how the visibility times for each visibility criterion relate to the different physical properties 
of each merger run, e.g., gas mass, viewing angle, etc.
  
Originally these parameters were envisaged to classify the morphologies of galaxies and hence a single parameter criterion will not necessarily work. Interacting 
spiral disks are a subset of the parameter-space occupied by late-types. \cite{CAS} and \cite{Lotz04} introduced several different criteria for the selection of merging 
systems in their respective parameter systems.
For optical data, \cite{CAS} define the following criterion:
\begin{equation}
A > 0.38, 
\label{eq:lcrit1}
\end{equation}
\noindent with some authors requiring A $>$ S as well. In general, highly asymmetric galaxies are candidate merging systems (Figure \ref{f:HI}).

\cite{Lotz04} added two different criteria using Gini and $M_{20}$:
\begin{equation}
G > -0.115 \times M_{20} + 0.384
\label{eq:lcrit2}
\end{equation}
and
\begin{equation}
G > -0.4 \times A + 0.66 ~ \rm or ~ A > 0.4.
\label{eq:lcrit3}
\end{equation}
\noindent The latter being a refinement of the Conselice et al criterion in equation \ref{eq:lcrit1}.

These criteria were developed for optical morphologies, typically observed Johnson-B or sdss-g. 
Typical optical spatial resolution ($\sim 2$") is a factor of a few higher than the typical spatial resolution of the \hi \ maps (6-12", depending on the survey). 
Therefore, in the third paper in this series \citep{Holwerda10c}, we defined several possible criteria specifically for the \hi \ perspective using the CAS-G/$M_{20}$-$G_M$ space of the WHISP survey \hi map sample.
We defined the Gini parameter of the second order moment, $G_M$ and a criterion that selected most interacting galaxies:
\begin{equation}
G_M > 0.6,
\label{eq:crit1}
\end{equation}
\noindent a criterion that seems to be corroborrated by Figure \ref{f:HI}.

Earlier in this series, we speculated that a combination of Asymmetry and $M_{20}$ could well be used to select interaction in \hi \ morphology in \citep{Holwerda10b}.
In  \cite{Holwerda10c}, we defined this criterion as:
\begin{equation}
A > -0.2 \times M_{20} + 0.25.
\label{eq:crit2}
\end{equation}

Finally, we also defined one based on Concentration and $M_{20}$, following the example of the \cite{Lotz04} criteria (eq. \ref{eq:lcrit2} and \ref{eq:lcrit3}):
\begin{equation}
C > -5 \times M_{20} + 3. 
\label{eq:crit3}
\end{equation}
In \citep{Holwerda10b}, we found that this last criterion selected to both the correct fraction of interacting galaxies as well as agree most often with the previous 
visual identifications in the case of individual WHISP galaxies.

We can now explore the timescales of each of these criteria, both those from the literature as well as those we determined in the WHISP sample subset.
The most reliable merger selection criterion would be the one that selects the right fraction of galaxies out of a given sample for a long timescale, with 
little contamination from isolated galaxies. Preferably, viewing angle or physical disk characteristics (e.g., gas mass) do not influence the timescale overly much.
A long selection time scale would subsequently ensure the more accurate estimate of the merger rate from a given volume of galaxies.

\begin{figure*}
\centering
\includegraphics[width=\textwidth]{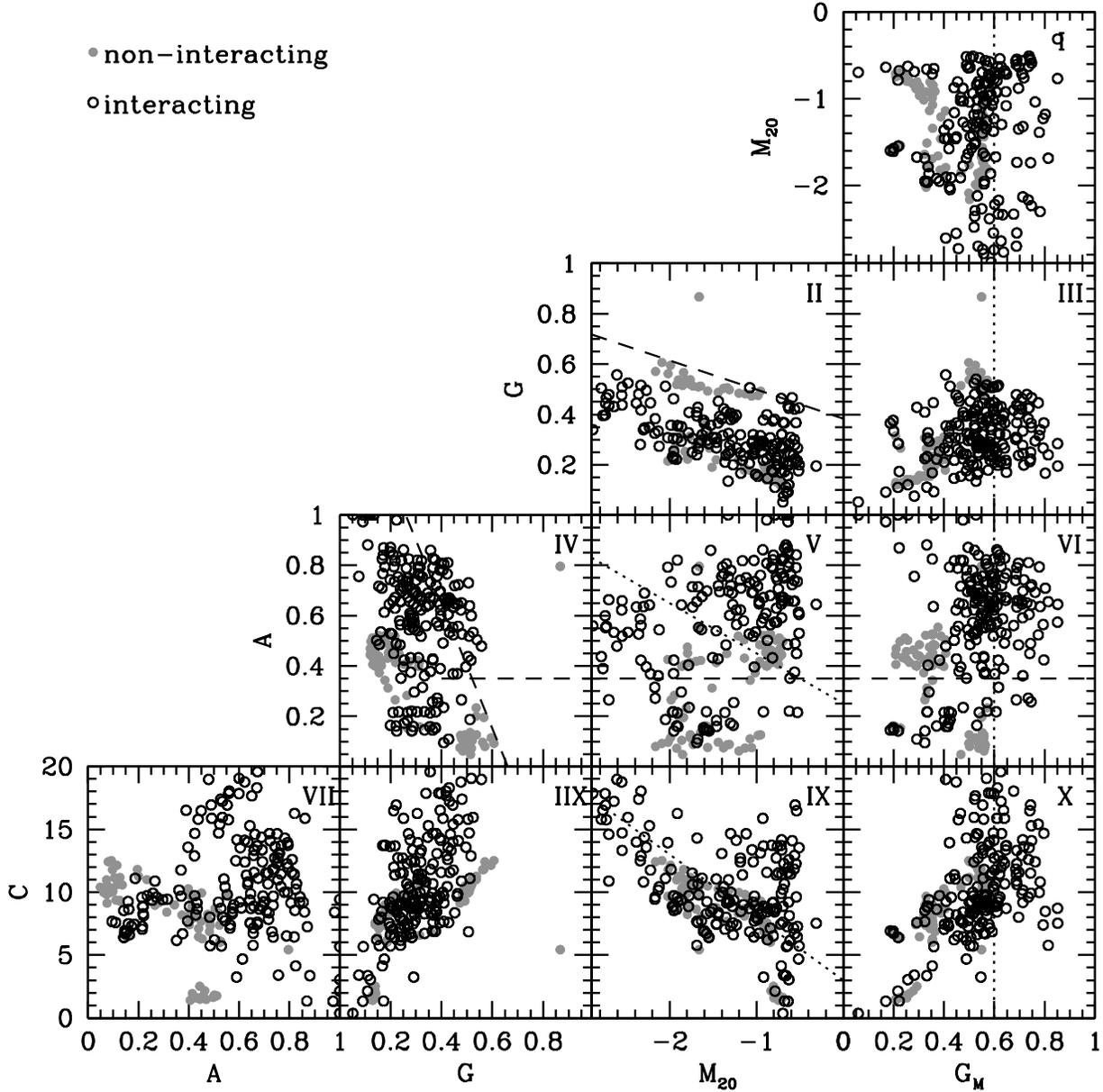}
\caption{\label{f:claudia_gas} The parameters of isolated disks (gray points) and mergers (black points) of all time-points in all simulations. The (optical) merger selection criteria from the literature (\protect\cite{CAS} and \protect\cite{Lotz04}) are marked with dashed lines in panel II (equation \protect\ref{eq:lcrit2}, below the line is a merger), panel IV (equations \protect\ref{eq:lcrit1} and \ref{eq:lcrit3}, above or to the right of the lines is a merger), and V and VI (equation \protect\ref{eq:lcrit1}, above the line is a merger). Our selection criteria from \protect\cite{Holwerda10c} are marked with dotted lines; the $G_M$ criterion in panels I, III, VI and X (equation \protect\ref{eq:crit1}, right of the line is a merger), the $A$-$M_{20}$ criterion in panel V (\protect\ref{eq:crit2}, above the line is a merger), and the C-$M_{20}$ criterion in panel IX (equation \protect\ref{eq:crit3}, above the line is a merger). Similar to Figure 4 in \protect\cite{Holwerda10c}. }
\end{figure*}

Figure \ref{f:claudia_gas} shows the relations between the morphological parameters for the \hi \ gas disks in the Cox et al. simulations, similar to 
Figure 4 in \cite{Holwerda10c}, from which we determined the selection criteria for \hi \ morphologies (eq. \ref{eq:crit1}-\ref{eq:crit3}). The literature and our 
merger selection criteria are marked with dashed and dotted lines respectively. The times an \hi \ disk is selected by the various selection criteria are listed 
in Table \ref{t:times}. 

\begin{figure*}
\centering
\includegraphics[width=0.32\textwidth]{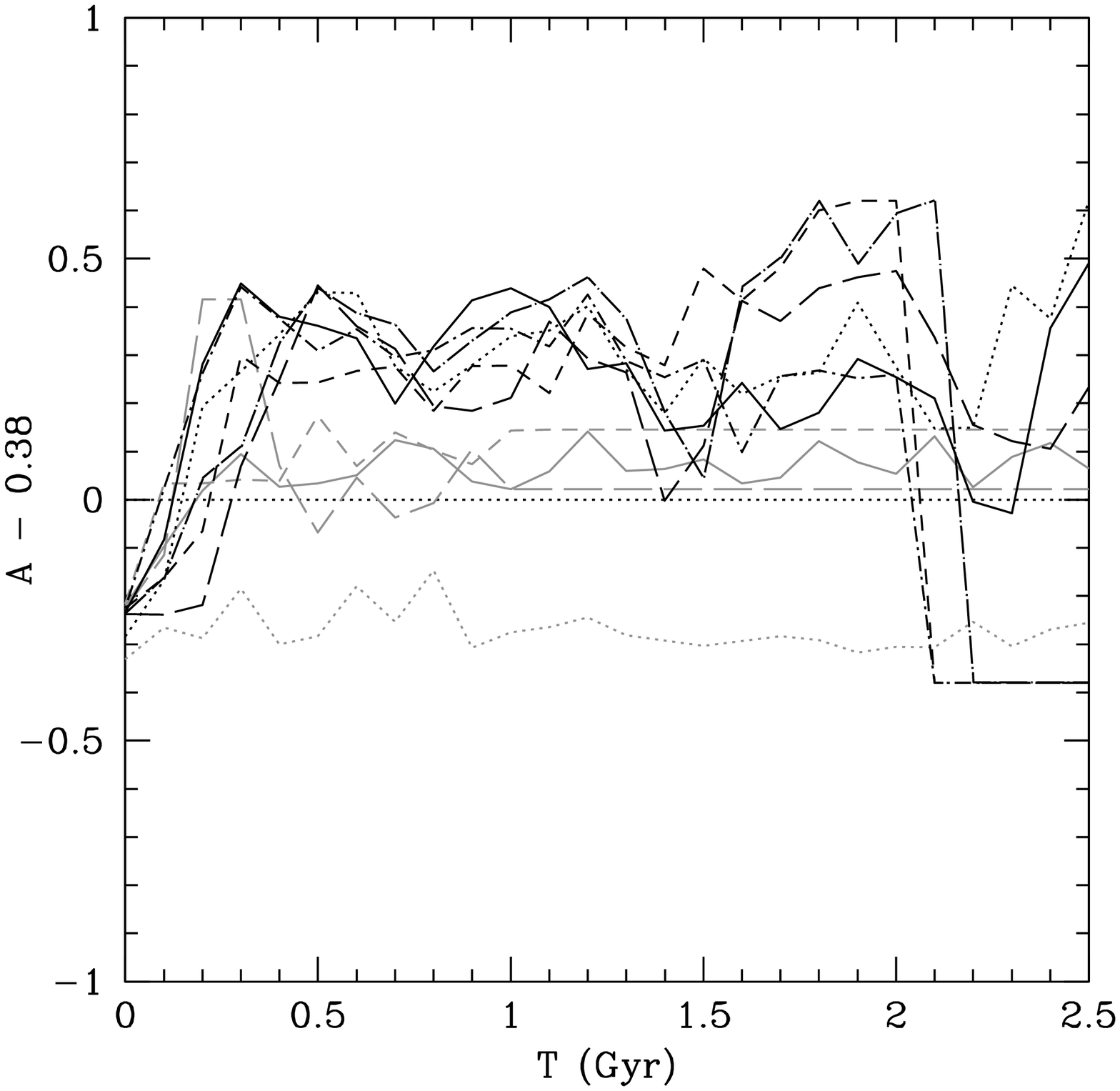}
\includegraphics[width=0.32\textwidth]{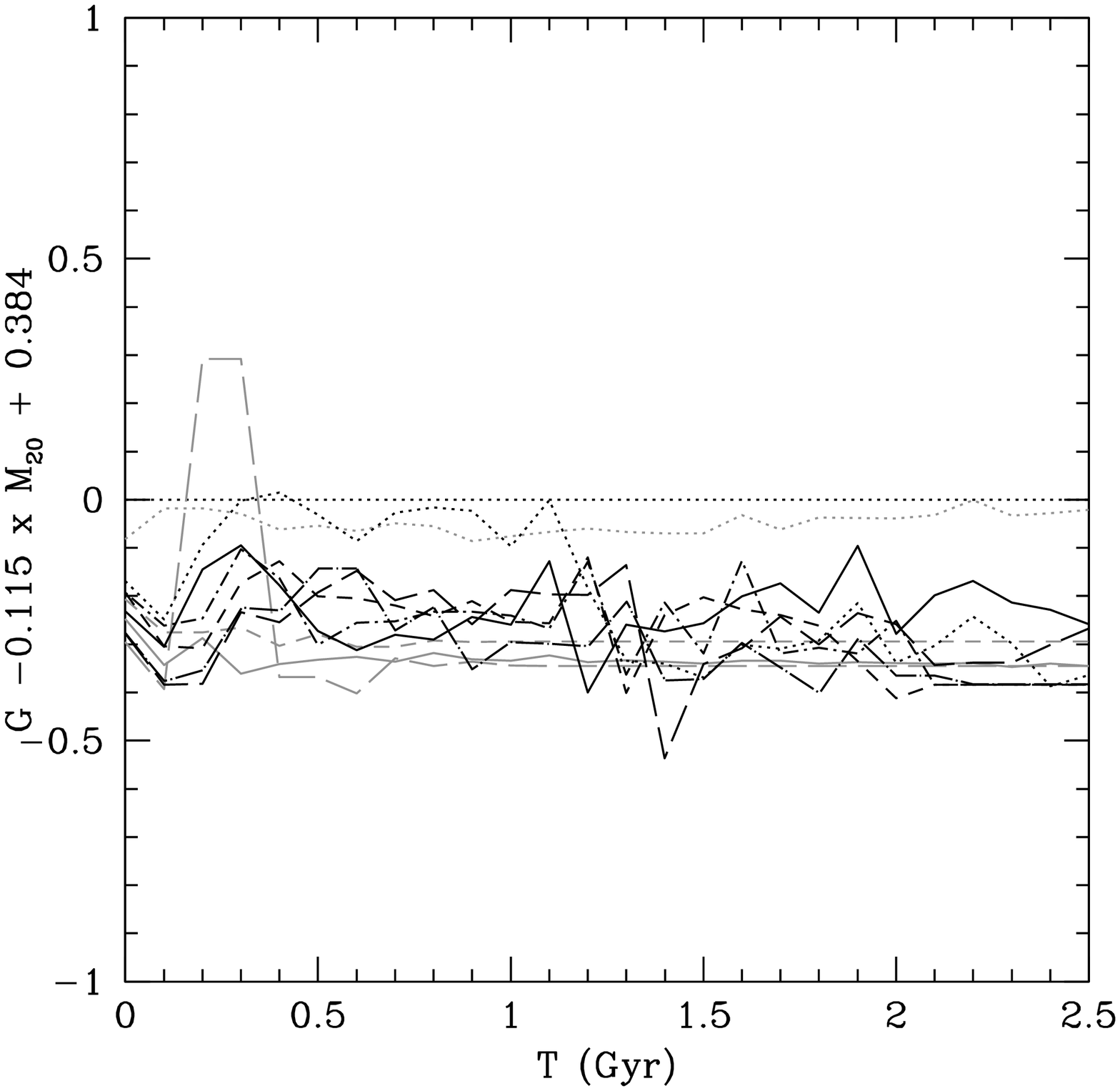}
\includegraphics[width=0.32\textwidth]{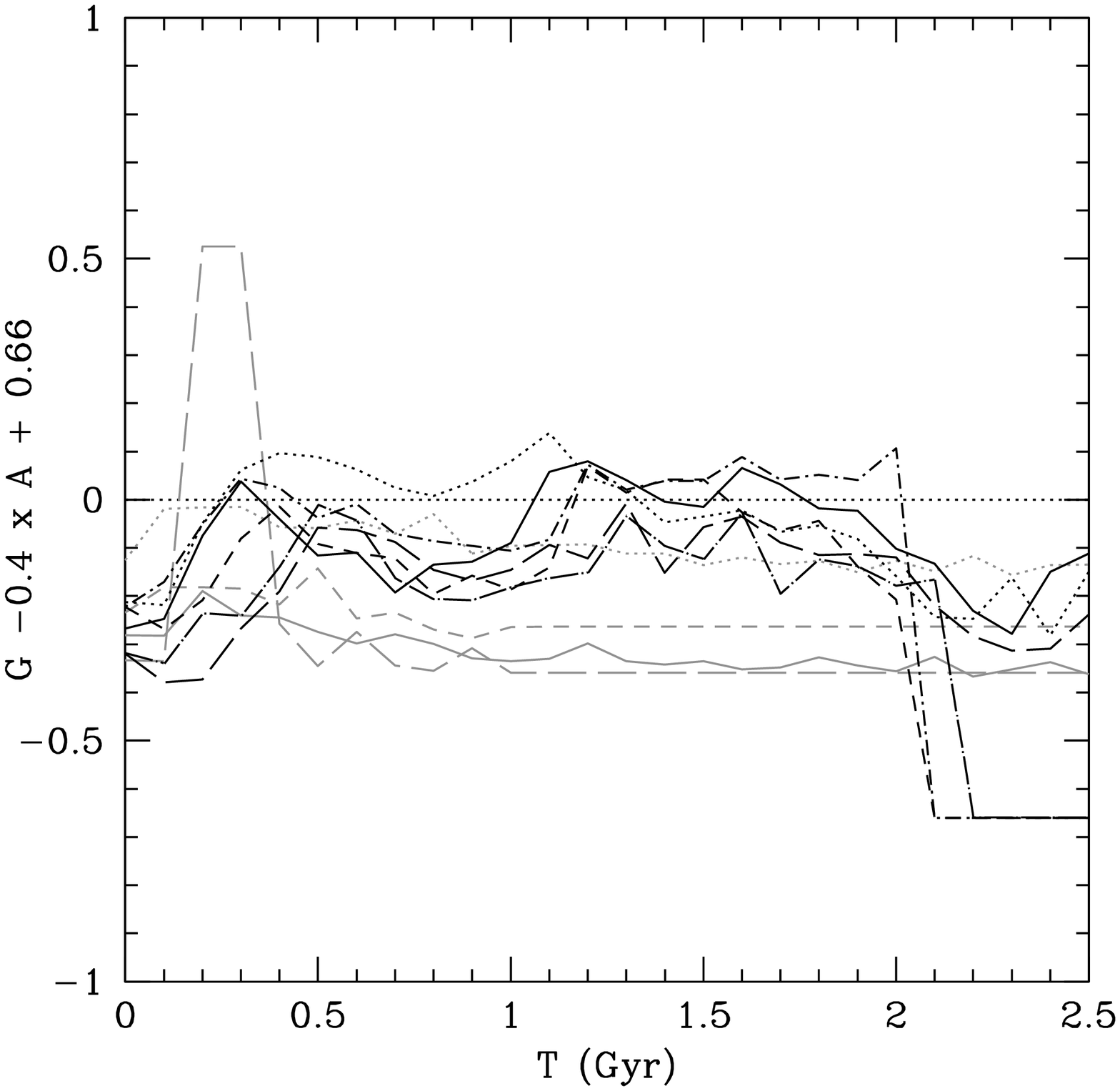}
\includegraphics[width=0.32\textwidth]{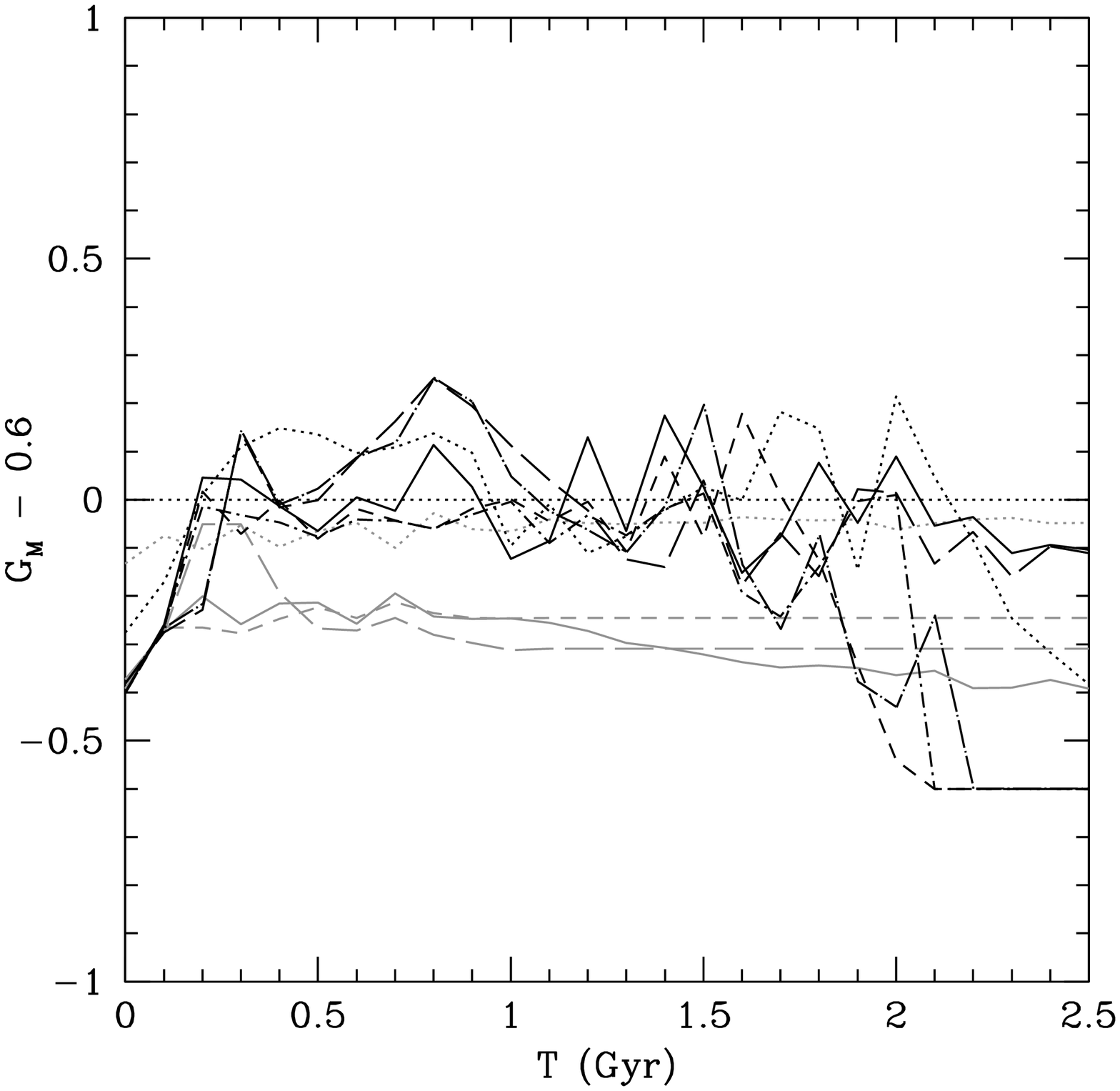}
\includegraphics[width=0.32\textwidth]{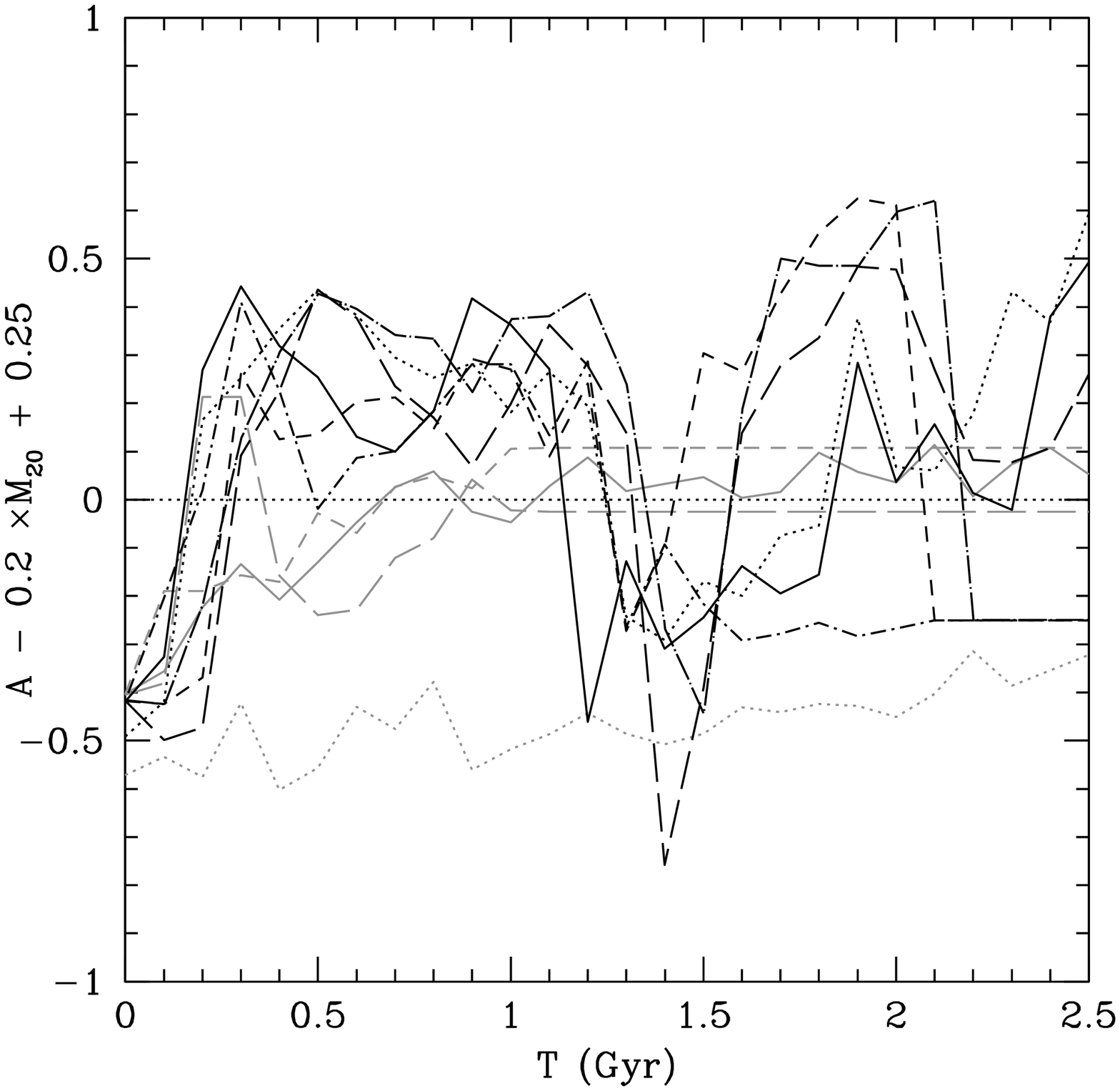}
\includegraphics[width=0.32\textwidth]{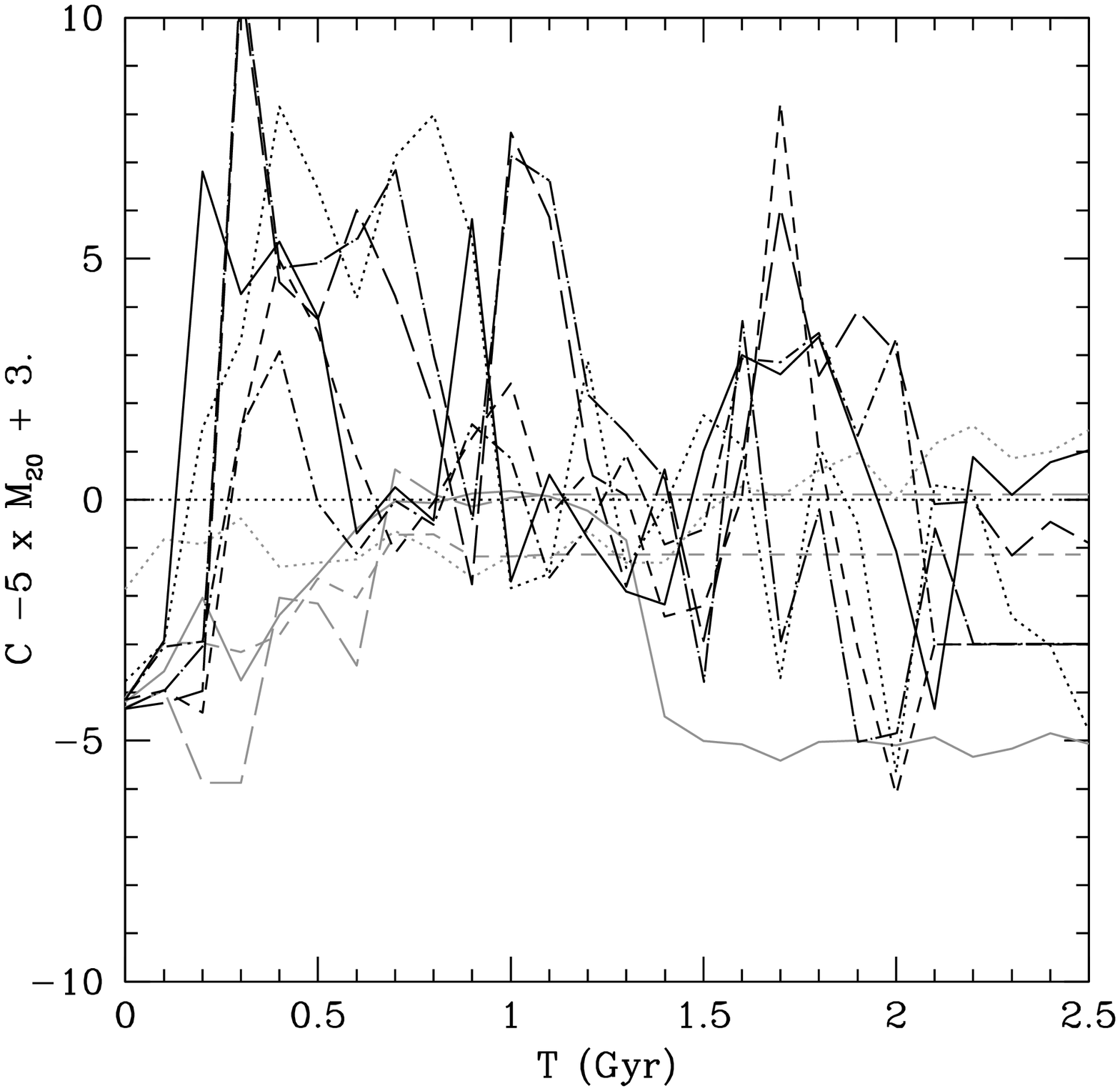}
\caption{\label{f:crit} Time evolution of the six criteria (equations \ref{eq:lcrit1}=-\ref{eq:crit3}) for different simulations of isolated disks (gray lines) and merging (black lines). A positive value indicates selection by the criterion as a merger. Different simulations are marked as in Figure \ref{f:HI}; nominal disks (solid lines), higher mass (short dashed lines), lower mass (long dashed lines), with different ISM treatment (high-mass; dot-short dash, low-mass; dot-long dash), and edge-on (dotted lines). Top three panels are the criteria from the literature (eq. \ref{eq:lcrit1}--\ref{eq:lcrit3}), bottom three panels are those defined in  \protect\cite{Holwerda10c} specifically for WHISP \hi \ maps.}
\end{figure*}

Figure \ref{f:crit} shows the six criteria (eq. \ref{eq:lcrit1}--\ref{eq:crit3}) as a function of time for all our simulations, both isolated and merging disks. If the y-axis values are positive, the merger would have been selected by this criterion. Immediately, it is apparent, that the three criteria that we defined in \cite{Holwerda10c} perform very well, separating merging galaxies from isolated disks. It also is clear that selection happens most often in the fist 1.5 Gyr of the merger simulations and that some of the criteria could be adjusted for the native resolution of these simulations. 

The selection by Asymmetry alone (eq. \ref{eq:lcrit1} or similar) does not translate well to \hi. Asymmetry alone selects isolated and merging galaxy almost 
equal times (Figure \ref{f:crit}, top left panel, Table \ref{t:times}).  This validates what we found for this criterion in the WHISP sample and our suspicion in \cite{Holwerda10b} that \hi \ Asymmetry is 
influenced by other factors than merging. From Figure \ref{f:crit}, one could conclude that a more strickt Asymmetry criterion would work better but this is only valid for the resolution of our simulations.
The Gini-$M_{20}$ criterion (eq. \ref{eq:lcrit2}) would need to be modified for the \hi \ perspective and performs poorly in the separation of mergers from isolated 
disks in its current definition (Figure \ref{f:crit}, top middle panel). The Gini-Asymmetry criterion (eq. \ref{eq:lcrit3}, Figure \ref{f:crit} top right panel) performs much better, very rarely selecting the isolated disk and selecting mergers on average for 0.44 Gyr out of the 2.5 Gyr the simulations ran.

However, these simulations ran on the maximum spatial resolution (150 pc sampling). The selection criteria perform a little poorer for lower resolution (Appendix B in the {\em electronic edition}). For example, the $G$--$A$ selection criterion (eq. \ref{eq:lcrit3}), still performs well for the WHISP survey observational parameters (Table \ref{t:WHISP}) but with a typical selection time 
scale of 0.15 Gyr. These merger visibility times are close to the ones typically quoted for these selection criteria applied to optical images of stellar disks \citep[e.g.,][]{Lotz10a,Lotz10b}.

Of the literature criteria, only the G-A one appears to translate directly from optical to \hi \ morphology. Our selection criteria (eq. \ref{eq:crit1}--\ref{eq:crit3}) were based on a 
WHISP subsample of \hi \ column density maps for which we had visual estimates of interaction. 
Our $G_M$ criterion (eq. \ref{eq:crit1}) cleanly select mergers for 0.69 Gyr of the 2.5 Gyr runtime (Figure \ref{f:crit}, bottom left panel). The selection becomes diluted, however, when the simulations are smoothed to the WHISP resolution 
(Table B.3 {\em online appendix}) or worse (e.g., VIVA, Table B.4 {\em online appendix}).
The $A$--$M_{20}$ selection criterion (eq. \ref{eq:crit2})  appeared to work extremely well in the WHISP data. Yet, in the simulations, it selects isolated disks more often, on average, 
than merging ones (Figure \ref{f:crit}, bottom middle panel). One can conceivable still use criteria like this one, provided the selected galaxy fraction is corrected for the contamination by isolated disks. Or the criterion is adjusted to the resolution of the data.
The Concentration-$M_{20}$ criterion (eq. \ref{eq:crit3}) performs better, selecting mergers for 1.19 Gyr versus isolated disks for 0.69 Gyr, but would also need a substantial 
correction for the isolated disk contaminations (Figure \ref{f:crit}, bottom right panel). We note that the main reason isolated disks are selected by this criterion are the low gas-mass and edge-on disk simulation and this criterion performs much better for the low inclination disks. Contamination increases again when the simulations are degraded to WHISP resolution 
(Table B.3 {\em in the online appendix}).

In \cite{Holwerda10c}, we found that the Concentration-$M_{20}$ criterion flagged not only the correct fraction of galaxies as merging but also agreed in many cases with 
the visual classifications on which galaxies were interacting. Here we find that the merger visibility time scale is in a similar range or even a little better than the selection criteria from the literature for optical images of disks (typically $\leq$ 1 Gyr). Thus, depending on the volume surveyed and the spatial resolution of the survey, an \hi  \ survey of disks can provide an estimate of the merger rate at least as accurate as any optical one using these parameters. 

\cite{Lotz10a} note that gas-rich mergers tend to be selected more often in the optical morphological selections, because gas rich mergers trigger more star-formation, which leaves brighter tidal features in the disks before full merger. A similar selection biases is pertinent for \hi \ morphology: more gas-rich disks will stand out in 21 cm. observations. Lowering or increasing the gas-mass of the disks, does influence the selection times, as does a different treatment of the ISM physics ($ism2$ or the Weniger simulation), increasing the variance in merger visibility times. 

The perspective or camera angle on the merger does influence the selection time for all the different criteria. Oddly, the edge-on perspective appears to occasionally raise 
the time a merger gets selected by some of the morphological parameter criteria. However, this can be expected to be the poorest viewing angle, regardless whether or not the morphology is determined in \hi \ or optical.

The changes in perspective, gas disk mass and ISM treatment for different simulations give an indication of the spread in merger visibility times in Table \ref{t:times}, which is substantial. Therefore, we can determine an accurate merger {\em fraction} for a sample of galaxies associated with a known volume, yet the merger rate will still be subject to some uncertainty due to the range in possible merger visibility times. With these average values in Table \ref{t:times} in hand, morphologically selected fraction of an \hi \ survey can now be converted to a merger rate with reasonable accuracy \citep[][]{Holwerda10e}.

\begin{table*}
\begin{tabular}{l l l l l l l l l }
Criterion		&  1			& 2						& 3				& 4			& 5					& 6 & 7 & 8 \\
\hline
\hline
isolated 		& & & & & & & & \\
\hline
face-on    		& 2.2		& 0		& 0		& 0		& 0.8		& 0.2		& 0		& 0.1\\
high-mass  	& 2.4		& 0		& 0		& 0		& 0.6		& 0		& 0		& 0 \\
low-mass   	& 2		& 0.1		& 0.1		& 0		& 2.2		& 1.7		& 0 		& 1.7\\
edge-on    	& 0		& 0    	& 0		& 0    	& 2.5    	& 0.9  	& 0  		& 0.9\\
\hline
mean       		& 1.63   	& 0   		& 0   		& 0   		& 1.53   	& 0.68     & 0 &  0.65\\
rms        		& 1.01  	& 0.007   	& 0   		& 0   		& 0.70   	& 0.48     & 0 &  0.51\\

\hline
\hline
 mergers			& & & & & & & & \\
\hline
face-on run 1	& 1.2		& 0	& 0.4		& 0.6	  	& 1.9		& 0.5         & 0.4		& 0.1\\
face-on run 2	& 2.1		& 0	& 0.5		& 0.9         & 0.9		& 1.5        	& 0.3		& 0.4\\
45$^\circ$                 	& 1.3		& 0	& 0.5		& 0.6		& 1.9		& 1.1	    	& 0.4		& 0.6\\
edge-on     	& 2.3		& 0	& 1		& 1.2		& 0.7		& 1.3	      	& 0.2		& 0.2\\
high-m      		& 1.7		& 0	& 0.3		& 0.4		& 0.9		& 1      	& 0.1		& 0     \\ 
low-m       		& 2.1		& 0	& 0		& 0.9       	& 0.4		& 1.4      	& 0		& 0      \\
ism2 high-m 	& 1.9		& 0  	& 1 		& 0.1		& 1.5		& 0.9 	& 0.1		& 0.5 \\
ism2 low-m  	& 1.9		& 0  	& 0 		& 0.7		& 0.8		& 1.1		& 0 		& 0 \\
\hline
mean        		& 1.9		& 0      & 0.44    & 0.69     & 1.01	   & 1.19     & 0.16	&  0.24\\
rms         		& 0.4		& 0     & 0.12    & 0.21      & 0.33	   & 0.20     & 0.02 &  0.05\\

\hline
\hline
 Weniger			& & & & & & & & \\
iso        		& 0.01  & 0	& 0.01 	& 0    	& 3  		& 0     	& 0     	& 0 \\
mm         		& 1.15  & 0.50	& 1.10	& 0.40	& 2.30  	& 0.65	& 0.10     & 0.15 \\
\hline
\hline

\end{tabular}
\begin{tablenotes}\footnotesize 
\item[1] 1. $A>0.4$ from \cite{CAS}
\item[2] 2. $G>-0.115 M_{20}+0.384$ from \cite{Lotz04}  	
\item[3] 3. $G>-0.4 A+0.66$  from \cite{Lotz04}  	
\item[4] 4. $G_M>0.6$ from \cite{Holwerda10c}.
\item[5] 5. $A<-0.2 M_{20}+0.25$  from \cite{Holwerda10c}.
\item[6] 6. $C>-5 M_{20}+3$  from \cite{Holwerda10c}.
\item[7] 7. The combination of $G_M>0.6$ and $A<-0.2 M_{20}+0.25$.
\item[8] 8. The combination of $A<-0.2 M_{20}+0.25$ and $C>-5 M_{20}+3$.
\end{tablenotes}
\caption{\label{t:times} The times in Gyr an \hi \ simulation from \protect\cite{Cox06a} is selected by the different criteria. Criteria are from \protect\cite{Holwerda10c}, and those adopted from the literature. }
\end{table*}%

\section{Conclusions}
\label{s:concl}

In this paper we obtained an understanding of the morphology of \hi \ disks during a major merger. Observationally, we have already found that the morphological parameters, currently in use to classify galaxies and identify mergers in optical and UV wavelengths can be applied very productively to \hi \ maps. Applying these to the cold gas and \hi \ maps based on N-body simulations, we conclude the following:

\begin{itemize}
\item [1.] The current merger simulations do an excellent job reproducing the stellar disks and its morphology as well as an accurate cold atomic hydrogen (\hi) map. 
Morphology of the \hi \ maps follows those seen in observations well both in appearance (Figure \ref{f:snap}) and quantified with CAS, G and $M_{20}$ (Figure \ref{f:claudia_gas}).
\item [2.] Asymmetry, $M_{20}$ and $G_M$ are good parameters to distinguish mergers from isolated disks (Figure \ref{f:HI} and \ref{f:weniger}) but a combination of morphological parameters often works better (Figure \ref{f:crit}).
\item [3.] Both the Weniger et al and the Cox et al. merger simulations show very similar behavior in morphology (Figure \ref{f:HI} and \ref{f:weniger}). Different implementations of the ISM physics need not result in dramatically different global \hi \ morphology.
\item [4.] Disk gas mass, orientation or the treatment of ISM physics do not substantially change the separation in parameter space of merging and isolated disks in these three parameters (Figure \ref{f:HI}).
\item [3.] Even the edge-on perspective on a merger produces a morphological signature in the \hi \ column density maps (Figure \ref{f:edgeon}), yet this remains the poorest viewing angle, often confusing merging and isolated disks. 
\item [4] Of the merger criteria from the literature, the one based on the Gini and Asymmetry performs well for the \hi \ perspective, selecting mergers for 0.4 Gyr out of a runtime of 2.5 Gyr. 
\item [5] The merger criteria defined in \cite{Holwerda10c} specifically for \hi \  show a range of merger selection time scales as well as some contamination from isolated disks (Figure \ref{f:claudia_gas} and Table \ref{t:times}). Of these three, the $G_M$ selection appears to be the cleanest selection with 0.69 Gyr visibility time. 
\item [6] The criteria we found to perform the best on the WHISP sample, $C>-5 M_{20}+3$, has a visibility time of $\sim$ 1 Gyr. but with some contamination by isolated disks, mostly low gas-mass and edge-on ones.
\end{itemize}

In the future, more detailed and comprehensive suites of simulated \hi \ disks will become available, which will allow for determinations of merger time scales for both dynamical and morphological signatures in \hi \ observations. Such simulations will become necessary when the large volume surveys on the SKA precursors commence. Notably, the all-sky \hi \ survey, that will result from the WALLABY project with the ASKAP radio telescope and the Northern \hi \ survey with the APERTIF instrument on WSRT combined, will produce hundreds of thousands of \hi \ data cubes, at similar resolutions as WHISP. Automated merger classification of this surveys will be the only feasible way to delineate samples, as well as determine a more accurate merger rate in the local Universe from an \hi \ perspective. 
Hopefully, this observational effort will be accompanied by an expanded suite of merger simulations to obtain a more accurate mean selection rate and possibly a spread due to difference in ISM, orientation and disk masses. Following the reasoning of \cite{Lotz10a,Lotz10b}, we will need to determine merger visibility time scales for the gas-rich and unequal mergers for which these upcoming surveys will be especially sensitive. 

Another possible future use of \hi \ morphological parameters is to combine them with at other wavelengths, e.g., optical. The different times at which the disk becomes disturbed --earlier in \hi, followed by optical-- could be  used to estimate in what phase any given merger is in.

\section*{Acknowledgments}

We acknowledge support from the National Research Foundation of South Africa. The work of B.W. Holwerda  and W.J.G. de Blok is based upon research supported by the South African Research Chairs Initiative of the Department of Science and Technology and the National Research Foundation. A. Bouchard acknowledges the financial support from the South African Square Kilometer Array Project. J. Weniger is supported by the University of Vienna in the frame of the Inititivkolleg (IK) 'The Cosmic Matter Circuit' I033-N and would like to thank Christian Theis and Stephan Harfst for fruitful discussions.

\end{document}